\documentclass{aa}

\usepackage{amsmath}
\usepackage{graphicx}
\usepackage{graphics}
\usepackage{epsfig}
\usepackage{longtable}
\usepackage{blindtext}
\usepackage{xcolor}

\usepackage{txfonts}
\usepackage[colorlinks=true,linkcolor=blue,citecolor=blue,urlcolor=blue]{hyperref}%
\newcommand{\ci}{C\,{\sc i}}
\newcommand{\hi}{H\,{\sc i}}
\newcommand{\kms}{\,km\,s$^{-1}$}

\begin{document}

\title{New constraints on the physical conditions in H$_2$-bearing GRB-host damped Lyman-$\alpha$ absorbers\thanks{Based on observations collected at the European Southern Observatory, Paranal, Chile, under the Stargate consortium with Program ID: 0102.D-0662.}}

\titlerunning{Physical conditions of H$_2$-bearing GRB-DLAs}

\author{
K.~E.~Heintz\inst{1},
J.~Bolmer\inst{2},
C.~Ledoux\inst{3},
P.~Noterdaeme\inst{4},
J.-K.~Krogager\inst{4},
J.~P.~U.~Fynbo\inst{5,6},
P.~Jakobsson\inst{1},
S.~Covino\inst{7},
V.~D'Elia\inst{8,9},
M.~De~Pasquale\inst{10},
D.~H.~Hartmann\inst{11},
L.~Izzo\inst{12},
J.~Japelj\inst{13},
D.~A.~Kann\inst{12},
L.~Kaper\inst{13},
P.~Petitjean\inst{4},
A.~Rossi\inst{14},
R.~Salvaterra\inst{15},
P.~Schady\inst{2,16},
J.~Selsing\inst{5,6},
R.~Starling\inst{17},
N.~R.~Tanvir\inst{17},
C.~C.~Th\"one\inst{12},
A.~de~Ugarte~Postigo\inst{12},
S.~D.~Vergani\inst{18,4},
D.~Watson\inst{5,6},
K.~Wiersema\inst{17,19}, and
T.~Zafar\inst{20}
}
\institute{
Centre for Astrophysics and Cosmology, Science Institute, University of Iceland, Dunhagi 5, 107 Reykjav\'ik, Iceland \\
\email{keh14@hi.is}
\and
Max-Planck-Institut für extraterrestrische Physik, Giessenbachstraße, 85748 Garching, Germany
\and
European Southern Observatory, Alonso de C\'ordova 3107, Vitacura, Casilla 19001, Santiago 19, Chile
\and
Institut d'Astrophysique de Paris, CNRS-SU, UMR7095, 98bis bd Arago, 75014 Paris, France
\and
Cosmic DAWN Center NBI/DTU-Space
\and
Niels Bohr Institute, University of Copenhagen, Lyngbyvej 2, 2100 Copenhagen \O, Denmark
\and
INAF -- Osservatorio Astronomico di Brera, Via E. Bianchi 46, I-23807 Merate (LC), Italy
\and
Space Science Data Center, SSDC, ASI, via del Politecnico snc, 00133 Roma, Italy
\and
INAF -- Osservatorio Astronomico di Roma, via Frascati 33, I-00040 Monteporzio Catone, Italy
\and
Department of Astronomy and Space Sciences, Istanbul University, Beyazit, 34119, Istanbul, Turkey
\and
Department of Physics and Astronomy, Clemson University, Clemson, SC29634-0978, USA
\and
Instituto de Astrof\'isica de Andaluc\'ia (IAA-CSIC), Glorieta de la Astronom\'ia s/n, 18008 Granada, Spain
\and
Anton Pannekoek Institute for Astronomy, University of Amsterdam, Science Park 904, 1098 XH Amsterdam, The Netherlands
\and
INAF -- Osservatorio di Astrofisica e Scienza dello Spazio, via Piero Gobetti 93/3, 40129 Bologna, Italy
\and
INAF -- IASF/Milano, via Corti 12, I-20133 Milano, Italy
\and
Department of Physics, University of Bath, Claverton Down, Bath, BA2 7AY, UK
\and
Department of Physics and Astronomy, University of Leicester, University Road, Leicester LE1 7RH, UK
\and
GEPI, Observatoire de Paris, PSL University, CNRS, Place Jules Janssen, 92190 Meudon
\and
Department of Physics, University of Warwick, Coventry CV4 7AL, UK
\and
Australian Astronomical Optics, Macquarie University, 105 Delhi Road, North Ryde, NSW 2113, Australia
}
\authorrunning{K.~E.~Heintz et al.}

\date{Received 2019; accepted 2019; published 2019}

\abstract{We report the detections of molecular hydrogen (H$_2$), vibrationally-excited H$_2$ (H$^*_2$), and neutral atomic carbon (\ci), an efficient tracer of molecular gas, in two new afterglow spectra of GRBs\,181020A ($z=2.938$) and 190114A ($z=3.376$), observed with X-shooter at the Very Large Telescope (VLT). Both host-galaxy absorption systems are characterized by strong damped Lyman-$\alpha$ absorbers (DLAs) and substantial amounts of molecular hydrogen with $\log N$(\hi, H$_2$) = $22.20\pm 0.05,~20.40\pm 0.04$ (GRB\,181020A) and $\log N$(\hi, H$_2$) = $22.15\pm 0.05,~19.44\pm 0.04$ (GRB\,190114A). The DLA metallicites, depletion levels and dust extinctions are within the typical regimes probed by GRBs with [Zn/H] = $-1.57\pm 0.06$, [Zn/Fe] = $0.67\pm 0.03$, and $A_V = 0.27\pm 0.02$\,mag (GRB\,181020A) and [Zn/H] = $-1.23\pm 0.07$, [Zn/Fe] = $1.06\pm 0.08$, and $A_V = 0.36\pm 0.02$\,mag (GRB\,190114A). In addition, we examine the molecular gas content of all known H$_2$-bearing GRB-DLAs and explore the physical conditions and characteristics required to simultaneously probe {\ci} and H$^*_2$. We confirm that H$_2$ is detected in all {\ci}- and H$^*_2$-bearing GRB absorption systems, but that these rarer features are not necessarily detected in all GRB H$_2$ absorbers. We find that a large molecular fraction of $f_{\rm H_2} \gtrsim 10^{-3}$ is required for {\ci} to be detected. The defining characteristic for H$^*_2$ to be present is less clear, though a large H$_2$ column density is an essential factor. We find that the observed line profiles of the molecular-gas tracers are kinematically \lq cold\rq, with small velocity offsets of $\delta v < 20$\kms~from the bulk of the neutral absorbing gas. We then derive the H$_2$ excitation temperatures of the molecular gas and find that they are relatively low with $T_{\rm ex} \approx 100 - 300$\,K, however, there could be evidence of warmer components populating the high-$J$ H$_2$ levels in GRBs\,181020A and 190114A. Finally, we demonstrate that even though the X-shooter GRB afterglow campaign has been successful in recovering several H$_2$-bearing GRB-host absorbers, this sample is still hampered by a significant dust bias excluding the most dust-obscured H$_2$ absorbers from identification. {\ci} and H$^*_2$ could open a potential route to identify molecular gas even in low-metallicity or highly dust-obscured bursts, though they are only efficient tracers for the most H$_2$-rich GRB-host absorption systems. }

\keywords{galaxies: ISM, high-redshift -- ISM: molecules -- dust, extinction -- gamma-ray bursts: general -- gamma-ray burst: individual: GRB\,181020A, GRB\,190114A}

\maketitle

\section{Introduction}     
\label{sec:intro}


Long-duration gamma-ray bursts (GRBs) are linked to the deaths of massive stars \citep[see e.g.][]{Woosley06}. These cosmological beacons originate at redshifts as high as $z \gtrsim 8$ \citep{Salvaterra09,Tanvir09}, and appear to be promising tracers of star formation, especially at high ($z \gtrsim 3$) redshifts \citep[e.g.][]{Greiner15,Perley16,Palmerio19}. GRBs are typically followed by a short-lived, multiwavelength afterglow emission \citep[e.g.][]{Meszaros06}, which, when bright enough, can serve as a powerful probe of the conditions in the star-forming regions and the interstellar medium (ISM) in their host galaxies \citep{Jakobsson04,Fynbo06,Prochaska07}. The absorption in GRB host galaxy lines-of-sight is typically found to be highly neutral-hydrogen-rich \citep{Vreeswijk04,Watson06,Fynbo09} and most of them are classified as damped Lyman-$\alpha$ absorbers \citep[DLAs;][]{Wolfe86}. These systems are similar to those previously observed in the spectra of bright quasars, which are produced by intervening galaxies in the line of sight. The DLAs provide the most effective and detailed probe of neutral gas at high redshifts \citep[see e.g.][for a review]{Wolfe05}. GRB host-galaxy absorbers are among the strongest DLAs, probing the central-most regions of their hosts, compared to typical quasar DLAs that are more likely to probe the outskirts of the intervening galaxies \citep{Fynbo09}. This makes GRB-DLAs the ideal probes of the ISM in high-redshift star-forming galaxies \citep[reaching $z \sim 8$, e.g.][]{Salvaterra15,Bolmer18,Tanvir18}.

Given their direct link to star-formation and the very high column densities of neutral
gas typically detected in GRB afterglow spectra \citep[e.g.][]{Jakobsson06}, it was anticipated that most GRB absorbers would show the presence of molecular hydrogen H$_2$ \citep{Galama01}. The observed low detection rate of H$_2$ was therefore initially a puzzle, indicating an apparent lack of molecular gas in GRB-host absorption systems \citep[e.g.][]{Tumlinson07}. The observed H$_2$ deficiency was attributed to the typical low metallicities of the GRB-host absorbers observed with high-resolution spectrographs \citep{Ledoux09}, or due to stronger UV radiation fields \citep{Whalen08,Chen09}. Since the absorption signatures of H$_2$ are the Lyman-Werner bands located bluewards of the Lyman-$\alpha$ line, the search for H$_2$ from the ground was also limited to $z \gtrsim 2$. Moreover, at $z\gtrsim 4$ the H$_2$ features becomes even more challenging to detect due to the increased Lyman forest line density. The first hint of molecular gas in GRB-host absorption systems came from the tentative detection of H$_2$ in GRB\,060206 \citep{Fynbo06}. However, it was not until the remarkable afterglow spectrum of GRB\,080607 that the first unequivocal detection of H$_2$ in a GRB-host absorber was reported \citep{Prochaska09}. The immense luminosity of this GRB \citep{Perley11} and the high H$_2$ column density made it possible to detect the absorption features from the UV Lyman-Werner bands, even in the low-resolution optical spectroscopy obtained of this GRB afterglow. Since then, six more H$_2$-bearing GRB absorbers have been securely detected \citep{Kruhler13,DElia14,Friis15,Bolmer19}. Except for GRB\,080607, all of these were observed with the more-sensitive, medium-resolution X-shooter spectrograph on the Very Large Telescope (VLT) as part of the extensive VLT/X-shooter GRB (XS-GRB) afterglow legacy survey \citep{Selsing19}. 

Star-formation is driven and regulated by the availability of dense gas, which is expected to be in molecular form in the ISM \citep{Mckee07,Kennicutt12}. Identifying and characterizing the molecular gas-phase is therefore vital to understand how stars are formed. At high redshifts, the presence of H$_2$ is most commonly inferred indirectly from other molecular gas tracers such as CO in emission-selected galaxy surveys \citep{Solomon05,Carilli13}, but its relation to H$_2$ at high-$z$ and in the low-metallicity regime is still uncertain \citep{Bolatto13}. Detecting the features from H$_2$, and other molecular gas species in absorption, therefore provides a unique window into the typical molecular gas content of high-$z$, star-forming galaxies. Recently, \cite{Heintz19a} also showed that neutral atomic carbon ({\ci}) could be used as a tracer of H$_2$ in GRB absorbers, suggesting that a relatively large fraction ($\sim 25\%$) of GRB sightlines intersect molecular clouds, also in the low-resolution spectroscopic GRB afterglow sample of \cite{Fynbo09}. Rarer molecules have also been detected in GRB afterglows, such as CH$^+$ in GRB\,140506A \citep{Fynbo14} and vibrationally-excited H$_2$ (H$^*_2$) in GRBs\,080607 and 120815A \citep{Sheffer09,Kruhler13}. Identifying H$_2$-bearing clouds from these alternative molecular gas tracers might prove to be even more effective, since they can be detected even in low-resolution spectroscopy, and in very dust-obscured GRB afterglows at lower redshifts. Line emission from CO has also been detected in a small number of GRB host galaxies \citep{Hatsukade14,Stanway15,Michalowski16,Michalowski18,Arabsalmani18}, providing an alternative way to establish the presence of molecular gas in the environments of GRBs.

Here we present the observations and detection of H$_2$ in the two host-galaxy absorption systems of GRB\,181020A at $z=2.938$ and GRB\,190114A at $z=3.376$. These two systems bring the total number of observed H$_2$-bearing GRB absorbers up to nine. 
Both afterglow spectra also show absorption features from {\ci} and exhibit the third and fourth known detections of H$^*_2$ in GRB-host absorption systems, respectively. The aim of this work is to explore the defining characteristics required for the H$_2$-bearing GRB absorbers to simultaneously probe {\ci} and H$^*_2$ and consequently quantify the use of the latter as alternative tracers of molecular-rich gas.

The paper is structured as follows. In Sect.~\ref{sec:sample}, we present the observations of the two new GRB optical afterglows and the compiled sample of GRB-host absorbers with detected molecules. The absorption-line abundance analysis is described in Sect.~\ref{sec:sel}, with specific focus on the identified molecular gas tracers. The results related to the defining characteristics of GRB-host absorption systems to show the presence of H$_2$, {\ci}, and H$^*_2$ is provided in Sect.~\ref{sec:res}. In Sect.~\ref{sec:disc}, we explore the physical conditions of the molecular gas and discuss the potential implications of a significant dust bias affecting the H$_2$-detection probability in GRB-host absorbers. Finally, in Sect.~\ref{sec:conc} we present the conclusions of our work. Throughout the paper, errors denote the $1\sigma$ confidence level (unless stated otherwise) and column densities are expressed in units of cm$^{-2}$. We assume a standard flat cosmology with $H_0 = 67.8$\,km\,s$^{-1}$\,Mpc$^{-1}$, $\Omega_m = 0.308$ and $\Omega_{\Lambda}=0.692$ \citep{Planck16}. Gas-phase abundances are expressed relative to the solar abundance values from \cite{Asplund09}, where [X/Y] = $\log (N(\mathrm{X})/N(\mathrm{Y})) - \log( N(\mathrm{X})_{\odot}/N(\mathrm{Y})_{\odot})$, following the recommendations by \citet{Lodders09}. Wavelengths are reported in vacuum.

\begin{figure*} 
	\centering
	\epsfig{file=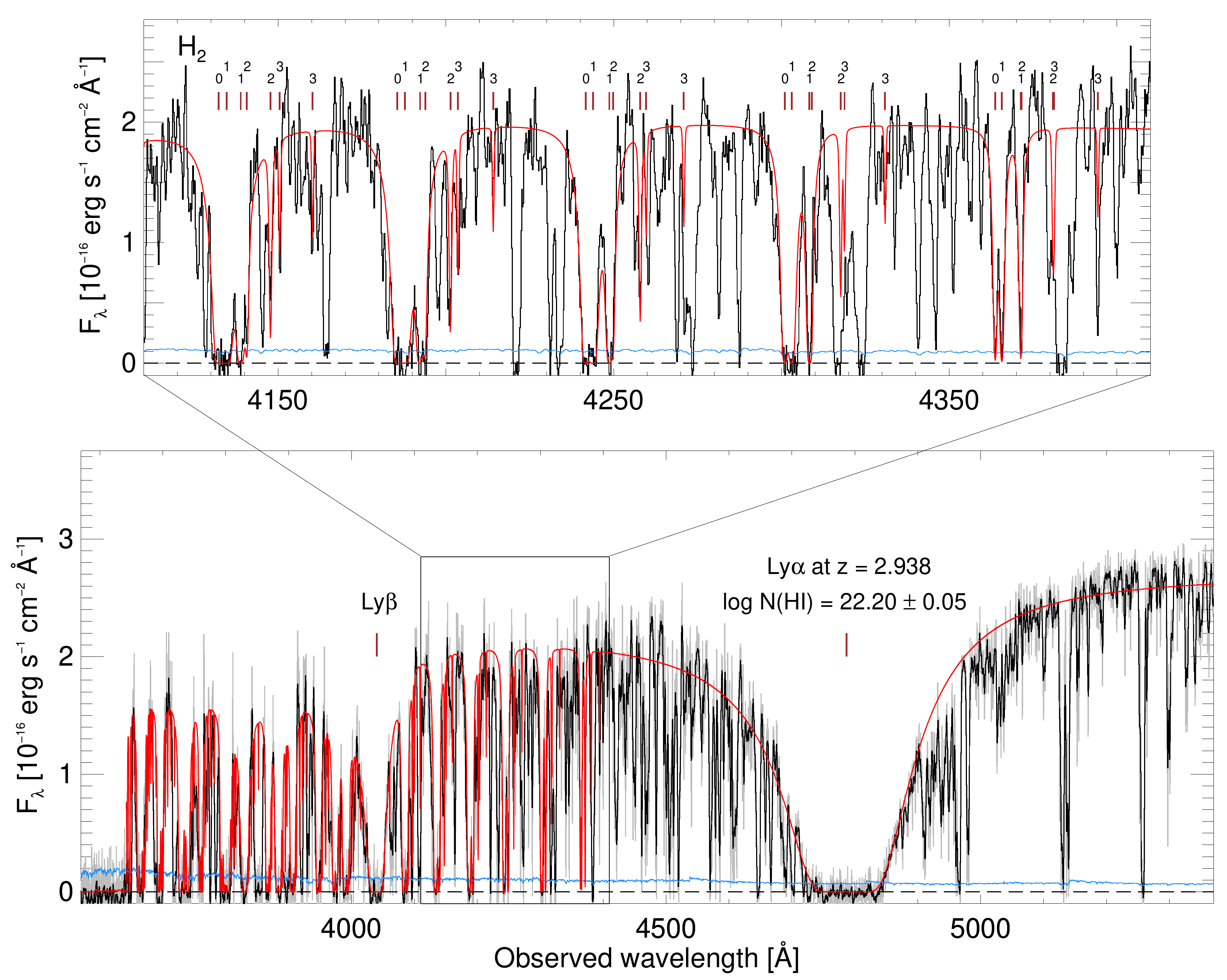,width=17cm}
	\caption{VLT/X-shooter UVB arm spectrum of GRB\,181020A showing absorption features from the {\hi} Ly$\alpha$, Ly$\beta$ and H$_2$ Lyman-Werner band line transitions. The raw spectrum is shown in grey with the associated error spectrum in blue. A binned version of the spectrum is overplotted in black. A synthetic spectrum of the best fit model to H\,{\sc i} and H$_2$ is shown as the red solid line and the Ly$\alpha$ and Ly$\beta$ absorption features are marked at the observed redshift of $z = 2.938$. The top panel shows a zoom-in of a subset of the most constraining Lyman-Werner band line transitions from molecular hydrogen. The redshifted wavelengths of the strongest rotational levels $J=0,1,2,3$ are marked above each line. 
	}
	\label{fig:grb181020a_spec}
\end{figure*} 

\section{Observations and sample description}    \label{sec:sample}

The GRBs\,181020A and 190114A were both detected by the Burst Alert Telescope \citep[BAT;][]{Barthelmy05} onboard the {\it Neil Gehrels Swift Observatory} \citep[{\textit{Swift} hereafter};][]{Gehrels04}, as reported by \citet{Moss18} and \cite{Laporte19}. We obtained optical to near-infrared afterglow spectra of both GRBs with the VLT/X-shooter instrument \citep{Vernet11}. The spectrum of GRB\,181020A was acquired (at the start of exposure) 5.7\,hrs after the GRB with an acquisition magnitude of $R\sim 17.3$\,mag \citep{Fynbo18b}. GRB\,190114A was observed only 15\,min after trigger with an acquisition magnitude of $R\sim 18.7$\,mag, using the rapid-response mode (RRM) \citep{deUgartePostigo19}. The optical to near-infrared spectra of both GRBs were taken simultaneously in the UVB, VIS and NIR arms of VLT/X-shooter with slit-widths of 1\farcs0, 0\farcs9, 0\farcs9, respectively. The observations were performed under excellent conditions with average seeing and airmasses of 1\farcs02, 1.25 (GRB\,181020A) and 0\farcs47, 1.29 (GRB\,190114A). Both spectra show very high signal-to-noise ratios (S/N), with S/N $\gtrsim 30$ at $\lambda_{\rm obs} = 6700$\,\AA~in the VIS arm spectra.

\begin{figure*} 
	\centering
	\epsfig{file=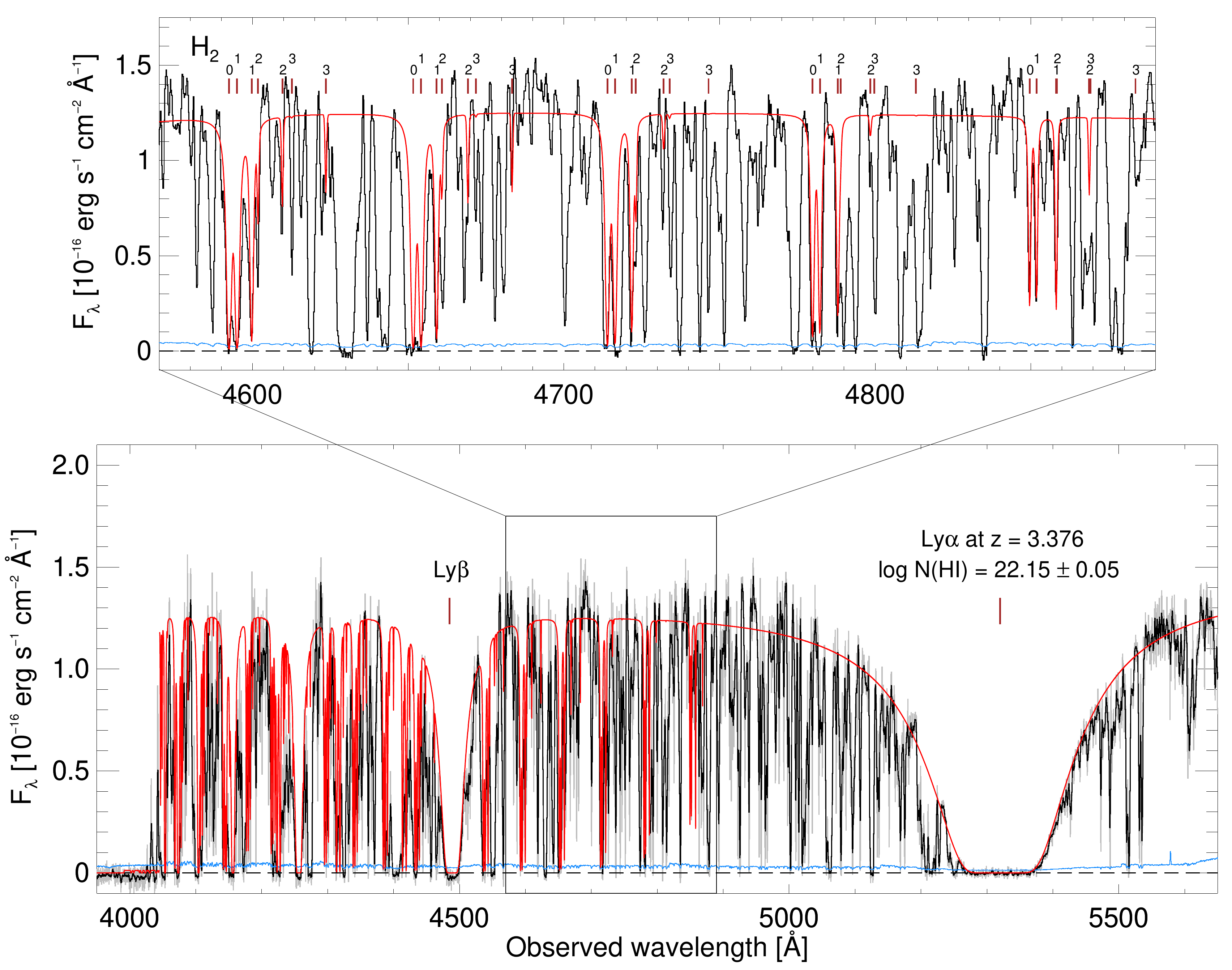,width=17cm}
	\caption{Same as Fig.~\ref{fig:grb181020a_spec} but for GRB\,190114A at $z = 3.376$. }%
	\label{fig:grb190114a_spec}
\end{figure*} 

We reduced the afterglow spectra in a similar manner to that described in \cite{Selsing19}. The only update is that the reductions are based on the most recent version {\tt v. 3.2.0} of the ESO X-shooter pipeline \citep{Modigliani10}. The final extracted 1D spectra were moved to the barycentric frame of the Solar system in the post-processing and corrected for Galactic extinction along the line-of-sight to the bursts using the dust maps of \citet{Schlegel98}, but with the updated values from \citet{Schlafly11}. We infer the delivered spectral resolution from the observed width of a set telluric absorption features in the VIS and NIR arms and rescale the nominal UVB arm resolution accordingly. To verify this, we also estimated the delivered resolution by fitting the spectral trace in the UVB arm with a Gaussian and derived the resolution from the full-width-at-half-maximum (FWHM) as FWHM = $2\sqrt{2\ln 2}\sigma$, which yielded consistent results.   

To complement the analysis of the molecular content and the physical conditions in GRBs\,181020A and 190114A, we compile a sample of bursts selected from the VLT/X-shooter GRB (XS-GRB) afterglow legacy survey \citep{Selsing19} and the Stargate (PI: N. R. Tanvir) public data. This sample constitutes all GRBs observed with VLT/X-shooter to date that directly or indirectly show the presence of molecular gas. We have extracted all afterglows with known detections of molecular hydrogen \citep[][]{Kruhler13,DElia14,Friis15,Bolmer19} and/or neutral atomic carbon \citep[][]{Zafar18a,Heintz19a}, which has been found to be a good tracer of molecular gas \citep[see e.g.][]{Srianand05,Noterdaeme18}. This collection is therefore not an unbiased representation of the GRB population as a whole, but rather a compilation of GRB afterglows where a determination of the relative gas content and the physical conditions in the molecular gas-phase is possible (see instead \citealt{Bolmer19} and \citealt{Heintz19a} for statistical analyses of the presence of H$_2$ and C\,{\sc i} in GRB afterglows). In total, we compiled ten XS-GRBs with H$_2$ and/or C\,\textsc{i} detected in absorption, listed in Table~\ref{tab:col}. Throughout the paper we will also compare our results to the only other GRB with a known detection of H$_2$ in absorption, GRB\,080607 \citep{Prochaska09}.

\section{Data analysis}    \label{sec:sel}

\subsection{Atomic and molecular hydrogen} \label{ssec:hih2}

In the high S/N afterglow spectra of GRBs 181020A and 190114A, we clearly detect the absorption features from H$_2$ bluewards of the broad Ly$\alpha$ absorption trough (see Figs.~\ref{fig:grb181020a_spec} and \ref{fig:grb190114a_spec}). We measure the column densities of atomic and molecular hydrogen by simultaneously fitting the absorption lines from H\,\textsc{i} and the H$_2$ Lyman-Werner bands following the same routine as described in \cite{Bolmer19}.
Here, the absorption lines are modelled with Voigt profiles and fitted simultaneously with the continuum flux of the GRB afterglow. The absorption lines are then convolved with the delivered spectral resolution in the UVB arm spectra of $\mathcal{R} = 6750$ (or $44.4$\kms, GRB\,181020A) and $\mathcal{R} = 7020$ (or $42.7$\kms, GRB\,190114A). The fitting routine is a custom-made \texttt{Python} module, based on the Markov Chain Monte Carlo (MCMC) Bayesian inference library \texttt{PyMC 2.3.7} \citep[see][for further details]{Bolmer19}. For GRB\,181020A we derive a total {\hi} column density of $\log N$(H\,\textsc{i}) = $22.20\pm 0.05$, consistent with \cite{Tanvir19}, at a redshift of $z=2.938$, and for GRB\,190114A we derive $\log N$(H\,\textsc{i}) = $22.15\pm 0.05$ at $z=3.376$. In both afterglow spectra, we detect all the rotational levels of H$_2$ up to $J=3$ (see Sect.~\ref{ssec:tex} for further discussion on the $J\ge 4$ levels).

\begin{table*}[!ht]
	\centering
	\begin{minipage}{\textwidth}
		\centering
		\caption{Sample properties of the H$_2$- and/or C\,{\sc i}-bearing XS-GRB absorbers.}
		\begin{tabular}{lcccccccc}
			\noalign{\smallskip} \hline \hline \noalign{\smallskip}
			GRB & $z_{\rm GRB}$ & $\log N$(H\,\textsc{i}) & $\log N$(H$_2$) & $\log f_{\mathrm{H}_2}$ & $\log N$(C\,\textsc{i})\tablefootmark{a} & $\log N$(CO)\tablefootmark{a} & [X/H] & $A_V$ \\
			& & (cm$^{-2}$) & (cm$^{-2}$) & & (cm$^{-2}$) & (cm$^{-2}$) && (mag) \\
			\noalign {\smallskip} \hline \noalign{\smallskip}
			120119A & 1.7288 & $22.44\pm 0.12$ & $\cdots$ & $\cdots$ & $\gtrsim 14.9$ & $<15.7$ & $-0.96\pm 0.28$ & $1.02\pm 0.11$  \\
			120327A & 2.8143 & $22.07\pm 0.01$ & $17.39\pm 0.13$ & $-4.38\pm 0.14$ & $<14.3$ & $<15.3$ & $-1.49\pm 0.03$ & $<0.03$ \\
			120815A & 2.3582 & $22.09\pm 0.01$ & $20.42\pm 0.08$ & $-1.39\pm 0.09$ & $14.24\pm 0.14$ & $<15.0$ & $-1.45\pm 0.03$ & $0.19\pm 0.04$  \\
			120909A & 3.9290 & $21.82\pm 0.02$ & $17.25\pm 0.23$ & $-4.27\pm 0.25$ & $<14.0$ & $<14.2$ & $-1.06\pm 0.12$ & $0.16\pm 0.04$  \\
			121024A & 2.3005 & $21.78\pm 0.02$ & $19.90\pm 0.17$ & $-1.59\pm 0.18$ & $13.91\pm 0.08$ & $<14.4$ & $-0.76\pm 0.06$ & $0.26\pm 0.07$ \\
			141109A & 2.9940 & $22.18\pm 0.02$ & $18.02\pm 0.12$& $-3.86\pm 0.14$  & $<14.7$ & $<15.9$ & $-1.63\pm 0.06$ & $0.11\pm 0.03$  \\
			150403A & 2.0571 & $21.73\pm 0.02$ & $19.90\pm 0.14$ & $-1.54\pm 0.15$ & $\gtrsim 14.3$ & $<14.9$ & $-1.04\pm 0.04$ & $<0.13$  \\
			180325A & 2.2496 & $22.30\pm 0.14$ & $\cdots$ & $\cdots$ & $\gtrsim 14.5$ & $<15.9$ & $>-0.96$ & $1.58\pm 0.12$ \\ 
			181020A & 2.9379 &$22.20\pm 0.05$ & $20.40\pm 0.04$ & $-1.51\pm 0.06$ & $13.98\pm 0.05$ & $<13.3$ & $-1.57\pm 0.06$ & $0.27\pm 0.02$  \\ 
			190114A & 3.3764 & $22.15\pm 0.05$ & $19.44\pm 0.04$ & $-2.40\pm 0.07$ & $13.54\pm 0.08$ & $<13.3$ & $-1.23\pm 0.07$ & $0.36\pm 0.02$ \\
			\noalign{\smallskip}\hline\noalign{\smallskip}
			080607 & 3.0363 & $22.70\pm 0.15$ & $21.20\pm 0.20$ & $-1.23\pm 0.24$ & $>15.1$\tablefootmark{b} & $16.5\pm 0.3$ & $> -0.2$\tablefootmark{c} & $2.58\pm 0.45$  \\ 
			\noalign{\smallskip} \hline \noalign{\smallskip}
		\end{tabular}
		\tablefoot{References for the measurements of the neutral and molecular gas-phase abundances and rest-frame $A_V$ can be found in the Appendix. \\
		\tablefoottext{a}{The measured $2\sigma$ upper or lower limits are provided for each GRB.} \\
		\tablefoottext{b}{The lower limit on the total C\,{\sc i} column density for GRB\,080607 is inferred from the rest-frame equivalent width of C\,{\sc i} (see Sect.~\ref{ssec:ci}).} \\
		\tablefoottext{c}{The lower limit on the metallicity for GRB\,080607 is derived from the [O/H] abundance following \cite{Prochaska09}.}
		}
		\label{tab:col}
	\end{minipage}
\end{table*}

To determine the H$_2$ abundances we tied the redshifts and $b$-parameters in the fit for all the detected rotational levels, and find a best fit assuming a single absorption component. For GRB\,181020A we measure column densities of $\log N$(H$_2$, $J=0,1,2,3$) = $20.14\pm 0.05$, $20.06\pm 0.03$, $18.38\pm 0.21$, and $18.05\pm 0.25$, and thus a total H$_2$ column density of $\log N$(H$_2$) = $20.40\pm 0.04$ with a broadening parameter of $b=3\pm 2$\,km\,s$^{-1}$. For GRB\,190114A we derive $\log N$(H$_2$, $J=0,1,2,3$) = $19.28\pm 0.05$, $18.90\pm 0.03$, $17.92\pm 0.02$, and $17.62\pm 0.25$ resulting in a total H$_2$ column density of $\log N$(H$_2$) = $19.44\pm 0.04$ with a broadening parameter of $b=2\pm 1$\,km\,s$^{-1}$. Since the lowest rotational levels ($J=0,1$) of H$_2$ are damped in both cases, the determination of the column density in these levels is not sensitive to $b$. Because these transitions dominate the H$_2$ content, the estimates of the total N(H$_2$) in both cases should be robust. While both fits are consistent with a single absorption component, we caution that at this resolution the observed line profiles might be comprised of a number of narrower features such that inferred H$_2$ abundances represent the integrated H$_2$ column density. 
Synthetic spectra of the best-fit models of H\,{\sc i} and H$_2$ in GRBs\,181020A and 190114A are shown in Figs.~\ref{fig:grb181020a_spec} and \ref{fig:grb190114a_spec}, overplotted on the UVB arm spectra.

For the remaining GRBs in our sample, column densities of atomic and molecular hydrogen have been derived previously in the literature. Throughout the paper, we report the column densities measured by \cite{Bolmer19} to be consistent within the sample, except for GRBs\,120119A and 180325A, where we adopt the derived H\,{\sc i} column densities from \cite{Wiseman17} and \cite{Zafar18a}, respectively (since they were not part of the statistical sample of \citealt{Bolmer19}). For a more detailed analysis of some of the individual systems, see the dedicated single-burst papers (e.g. for GRB\,120327A: \citealt{DElia14}; GRB\,120815A: \citealt{Kruhler13}; and GRB\,121024A: \citealt{Friis15}).

\subsection{Gas-phase abundances and dust extinction}

In addition to the {\hi} and H$_2$ transition lines we detect a wealth of low-ionization metal absorption features in the afterglow spectra of GRBs\,181020A and 190114A. To determine the gas-phase abundances, we again fit a range of Voigt profiles to a set of carefully-selected absorption lines, free of tellurics or unrelated blends. The host absorber towards GRB\,181020A shows a complex velocity structure with five identified strong absorption components (see Fig.~\ref{afig:181020a_met} for a few examples showing this structure). The absorption line profiles in the host absorber of GRB\,190114A show a simpler velocity structure, with one dominant component and an additional weaker component at $\delta v = -40$\kms{} (see Fig.~\ref{afig:190114a_met}). We constrain the column densities from the weakest transitions of each element by fixing the velocity structure to that observed in the strongest line complexes. 

For GRB\,181020A we derive column densities of $\log N$(Fe) = $15.47\pm 0.01$, $\log N$(Zn) = $13.19\pm 0.03$, and $\log N$(Cr) = $14.16\pm 0.02$, resulting in a gas-phase metallicity of [Zn/H] = $-1.57 \pm 0.06$ and dust-depletion [Zn/Fe] = $0.67\pm 0.03$. 
Following \citet{DeCia16} we compute a dust-corrected metallicity, [M/H] = [X/H] - $\delta_X$ (where $\delta_X$ is inferred from the iron-to-zinc depletion), of [M/H] = $-1.39\pm 0.05$. For GRB\,190114A we measure gas-phase abundances of $\log N$(Fe) = $15.37\pm 0.04$, $\log N$(Zn) = $13.48\pm 0.04$, and $\log N$(Cr) = $13.95\pm 0.05$, resulting in a metallicity of [Zn/H] = $-1.23 \pm 0.07$ and dust-depletion [Zn/Fe] = $1.06\pm 0.08$. This yields a dust-corrected metallicity of [M/H] = $-0.94\pm 0.06$. The gas-phase abundances of GRBs\,181020A and 190114A, together with literature values for the other GRBs in our sample, are summarized in Table~\ref{tab:col}. Again, we adopt the values derived by \citet{Bolmer19} for the majority of the sample, except for GRBs\,120119A and 180325A (see the Appendix for details).

The visual extinction $A_V$ along the line-of-sight to both GRBs were measured following the same approach as in \cite{Heintz19a}. Briefly, this assumes a simple underlying power-law shape of the afterglow spectrum, with a wavelength-dependent extinction coefficient $A_{\lambda}$ imposed as $F_{\rm obs} = F_{\lambda} \times 10^{-0.4 A_{\lambda}}$ where $F_{\lambda} = \lambda^{-\beta}$. We then fit the underlying power-law slope and extinction coefficient simultaneously. Using the extinction-curve parametrization from \citet{Gordon03}, we find a best fit with an SMC-like extinction curve in both GRB sightlines and measure $A_V = 0.27\pm 0.02$~mag (for GRB\,181020A, see Fig.~\ref{afig:181020a_ext}) and $A_V = 0.36\pm 0.02$~mag (for GRB\,190114A, see Fig.~\ref{afig:190114a_ext}). We do not find any indication of the 2175\,\AA~dust extinction bump in either of the bursts, so we derive upper limits on the bump strength, $A_{\rm bump} = \pi c_3 /(2\,\gamma\,R_V)\times A_V$, of $A_{\rm bump} < 0.07$\,mag (GRB\,181020A) and $A_{\rm bump} < 0.09$\,mag (GRB\,190114A) at $3\sigma$. The measured visual extinction along the line-of-sight toward GRBs\,181020A and 190114A, together with literature values for the other GRBs in our sample, are again summarized in Table~\ref{tab:col}.

\begin{figure} 
	\centering
	\epsfig{file=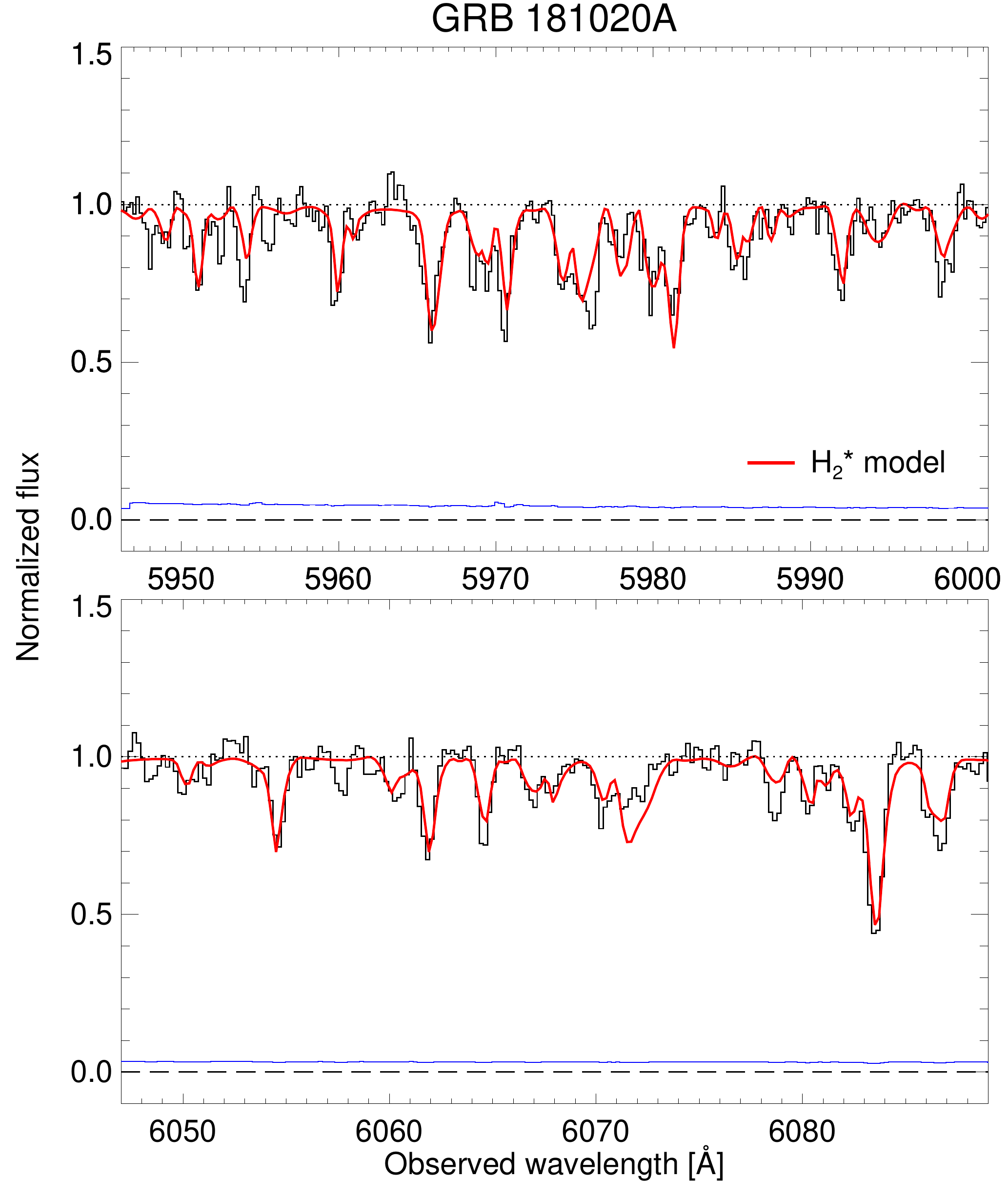,width=8.5cm}
	\caption{Normalized VLT/X-shooter VIS arm spectrum of GRB\,181020A showing regions encompassing a subset of the strongest H$^*_2$ lines. The black solid lines show the data, the blue lines the associated error and the red lines the best-fit model. }
	\label{fig:H2vib181020A}
\end{figure}

\subsection{Vibrationally-excited molecular hydrogen}

After clearly establishing the presence of H$_2$ in both the afterglow spectra of GRBs\,181020A and 190114A, we searched for the so-far rarely detected absorption features from vibrationally-excited H$_2$ (H$^*_2$) \citep[see][]{Bolmerth}. The vibrationally-excited levels of H$_2$ are expected to be populated by UV pumping from the GRB afterglow \citep{Draine00}, but to date they have only been securely detected in two afterglow spectra, those of GRB\,080607 \citep{Sheffer09} and GRB\,120815A \citep{Kruhler13}. We performed the search by using the synthetic spectrum from \citet{Draine02}, downgraded to the resolution of the given arm (in both cases the VIS arm, with $\mathcal{R} \sim 11\,000$). In the fit, we included any intervening metal lines and matched the model to the whole spectrum redwards of Ly$\alpha$ up until 1650\,{\AA} (rest frame). We use PyMC (as described in Sect.~\ref{ssec:hih2}) to sample the posteriors of the optical depth $\tau$ and the redshift of the H$^*_2$ absorption lines, as well as the continuum flux.

\begin{figure} 
	\centering
	\epsfig{file=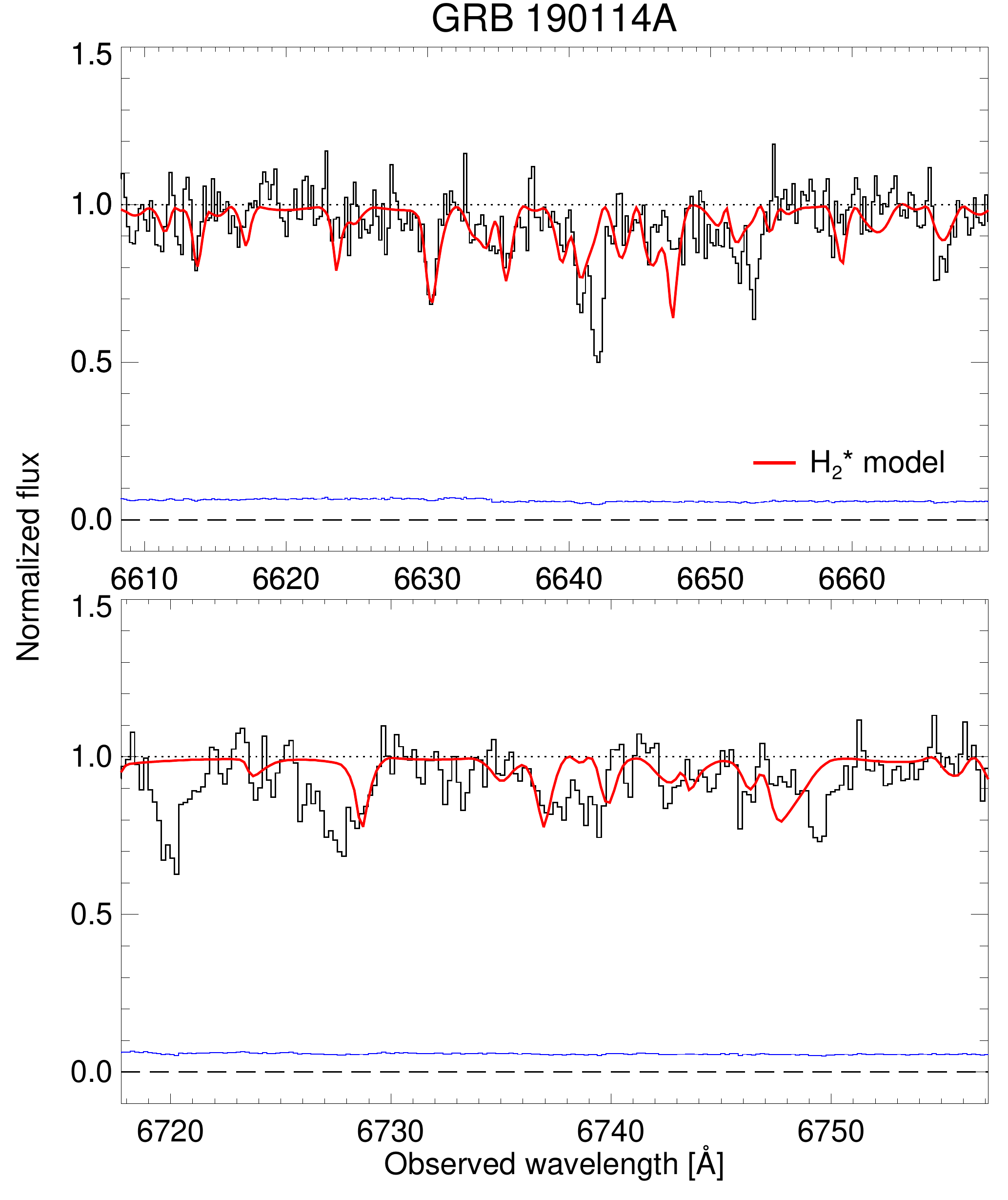,width=8.5cm}
	\caption{Same as Fig.~\ref{fig:H2vib181020A} but for GRB\,190114A.} 
	\label{fig:H2vib190114A}
\end{figure}

We clearly detect H$^*_2$ in both afterglow spectra of GRBs\,181020A and 190114A. Even including the uncertainty on the continuum flux due to the wealth of absorption features from H$^*_2$, we consider the detections highly significant due to the overall excellent match with the data and fit to several strong individual lines. For GRB\,181020A we derive a column density of $\log N(\mathrm{H}^*_2)=16.28\pm0.05$, with the best-fit model shown in Fig.~\ref{fig:H2vib181020A}. For GRB\,190114A, we measure $\log N(\mathrm{H}^*_2)=16.13\pm0.13$, with the best-fit model shown in Fig.~\ref{fig:H2vib190114A}. We caution that small deviations of the spectrum from the model are expected due to the different initial conditions, such as the luminosity of the GRB afterglow, dust content and shielding, and distance to the absorbing cloud \citep{Kruhler13}. A detailed modelling of the lines will be provided in a follow-up paper to constrain the origin of H$^*_2$ (Bolmer et al. in preparation). The H$^*_2$ column densities of GRBs\,181020A and 190114A are of the same order as the one observed in GRB\,120815A, all being roughly an order of magnitude lower than what was observed in GRB\,080607 \citep{Sheffer09}. For the other GRBs in our sample, we are able to place upper limits on the H$^*_2$ column density typically $5-10$ times lower than those observed in GRBs\,120815A, 181020A and 190114A using the same routine \citep[see also][]{Bolmer19}. 

\begin{figure} 
	\centering
	\epsfig{file=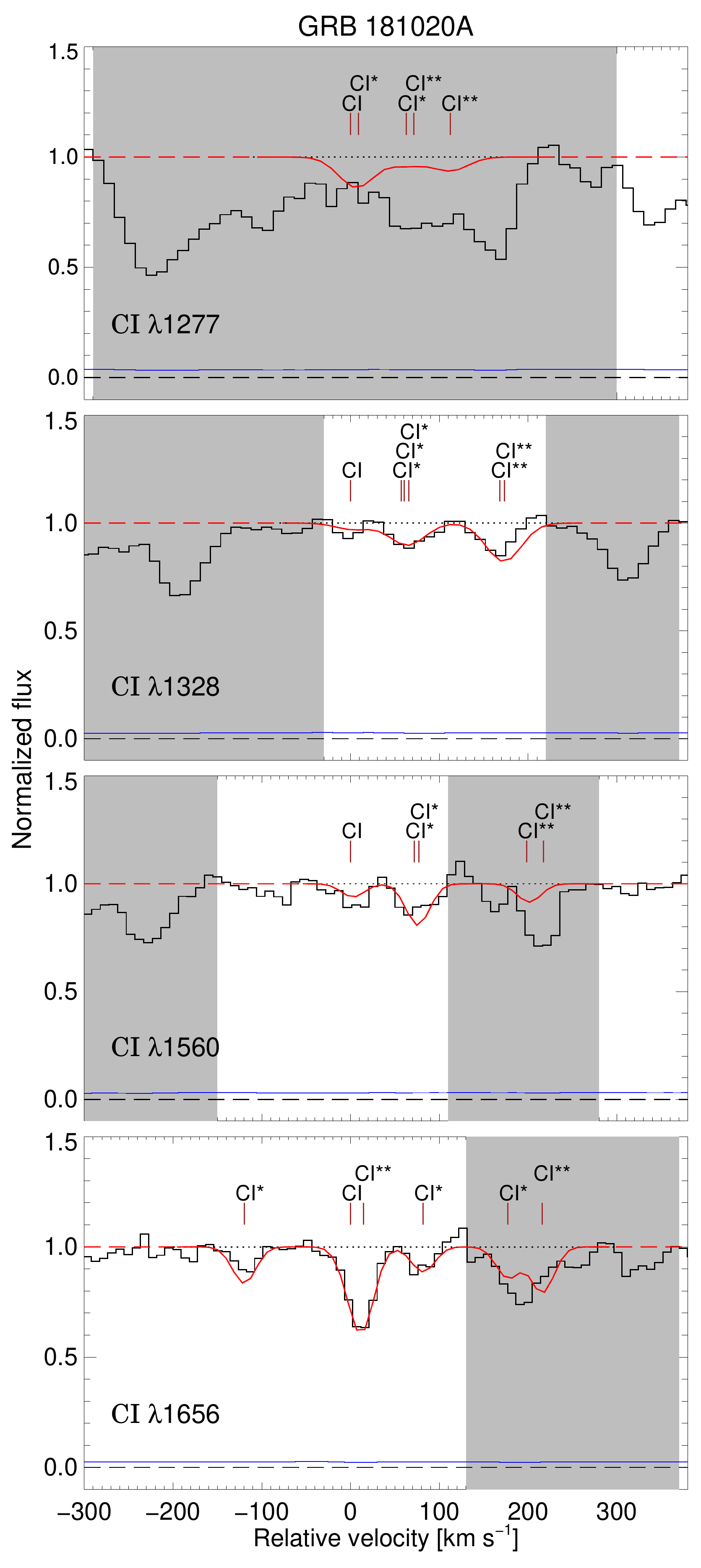,width=8.5cm}
	\caption{Normalized VLT/X-shooter VIS arm spectrum of GRB\,181020A in velocity space, centred on the ground-state transition of {\ci} at $z=2.93786$. Again, the black solid lines show the data, the blue lines the associated error and the red lines the best-fit model. The C\,\textsc{i} ground-state and excited line transitions are marked above each of the absorption profiles. Gray shaded regions were ignored in the fit.}
	\label{fig:ci181020a}
\end{figure} 

\begin{figure} 
	\centering
	\epsfig{file=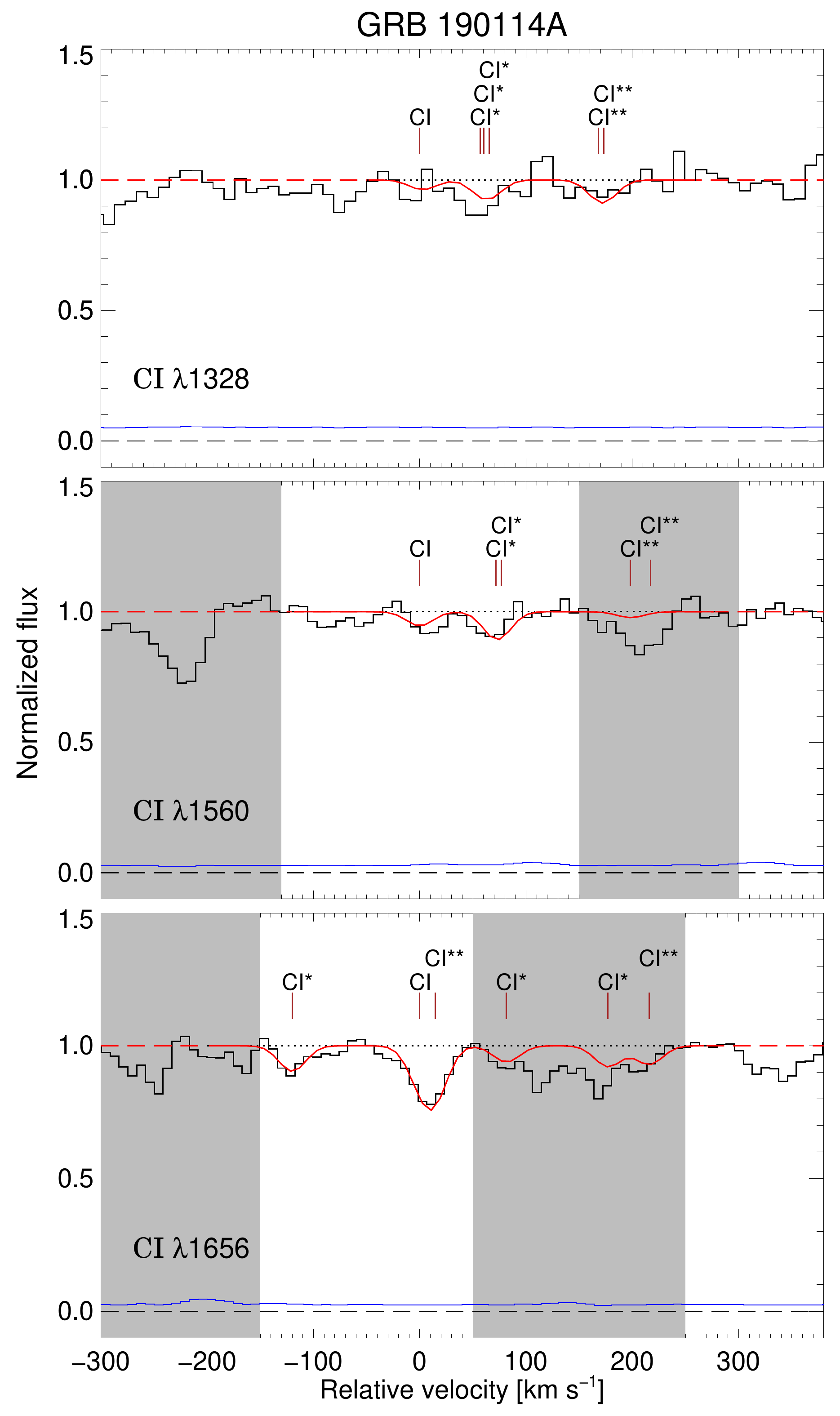,width=8.5cm}
	\caption{Same as Fig.~\ref{fig:ci181020a} but for GRB\,190114A and with a fixed value of $b=5$\kms, centred on $z=3.37638$.}
	\label{fig:ci190114a_b5}
\end{figure} 

\subsection{Neutral atomic carbon} \label{ssec:ci}

In addition to H$_2$ and H$^*_2$, we also detect absorption features from {\ci} in both afterglow spectra of GRBs\,181020A and 190114A, which has been found to be an efficient tracer of molecule-rich gas. Recently, \citet{Heintz19a} surveyed {\ci} in a large sample of GRB afterglows observed both with low- and medium-resolution spectrographs (including some of the XS-GRBs in the sample presented here). To be consistent with the part of the sample observed with low-resolution spectrographs (where meaningful column densities cannot be derived) and with the survey for C\,\textsc{i} in high-$z$ quasar absorbers \citep{Ledoux15}, only the total C\,\textsc{i} equivalent widths were measured in that study. In this work, we attempt to derive the C\,{\sc i} column densities for all the XS-GRBs, including the sample studied in \citet{Heintz19a}, by fitting a set of Voigt profiles to the relevant transitions. For this, we used the \texttt{Python} module \texttt{VoigtFit}\footnote{\url{https://github.com/jkrogager/VoigtFit}} \citep{Krogager18a}, where the absorption line profiles again have been convolved with the delivered instrumental resolution. 

The three fine-structure levels ($J = {0, 1, 2}$) of neutral carbon's ground-state triplet, here denoted C\,\textsc{i}, C\,\textsc{i}*, C\,\textsc{i}**, respectively, are all resolved in the VLT/X-shooter spectra \citep[see also e.g.][]{Krogager16,Ranjan18}. We simultaneously fitted the three fine-structure levels, assuming a single component and by tying the Doppler parameters and redshifts. This is based on the assumption that the excited fine-structure levels (C\,\textsc{i}* and C\,\textsc{i}**) share the same physical origin as the ground level (C\,\textsc{i}) and therefore follow the same kinematic structure.  

For GRB\,181020A we derive column densities of $\log N$(\ci, \ci*, \ci**$)=12.82\pm0.13$, $13.45\pm0.05$, and $13.78\pm0.07$, and thus a total {\ci} column density of $\log N$(\ci$)=13.98\pm0.05$ with a best-fit broadening parameter of $b=3.9\pm0.8$\,km\,s$^{-1}$. The best-fit Voigt profiles are shown in Fig.~\ref{fig:ci181020a}. We only detect a single absorption component across the four line complexes so we fixed this in the fit and masked out any unrelated or blended features. The fit was only constrained by the C\,\textsc{i}\,$\lambda\lambda\lambda$\,1328,1560,1656 transition lines, since the C\,\textsc{i}\,$\lambda$\,1277 line is significantly blended with tellurics and unrelated absorption features. For GRB\,190114A we compute relative {\ci} abundances of $\log N$(\ci, \ci*, \ci**$)=12.79\pm 0.15$, $13.12\pm0.15$, and $13.19\pm0.11$, and thus a total {\ci} column density of $\log N$(\ci$)=13.54\pm0.08$. In this case, we fixed the broadening parameter to $b=5$\kms, since the fit could not converge on a realistic $b$-value due to significant blending of several of the lines. The best-fit model with fixed $b=5$\kms~is shown in Fig.~\ref{fig:ci190114a_b5}. It was only possible to perform the fit on the C\,\textsc{i}\,$\lambda\lambda\lambda$\,1328,1560,1656 line transitions, since the C\,\textsc{i}\,$\lambda$\,1277 line is located in the overlap region between the UVB and VIS arm. We only detect a single absorption component across the three line complexes in this case as well, so we fixed this in the fit and masked out any unrelated or blended features. The line profiles seem to exclude values of $b\gtrsim 5$\kms~and $b\lesssim 3$\kms, both when considering the line widths and the relative optical depths.

For the other GRBs in our sample where C\,{\sc i} is detected, the derived column densities are listed in Table~\ref{tab:col}. Here, we also provide upper limits for the H$_2$-bearing GRB absorbers which show non-detections of C\,\textsc{i} assuming $b=2$\kms~to be consistent with the limits derived for the abundance of CO \citep{Bolmer19}. In the Appendix, a more detailed description of the fit performed for each individual GRB is given, together with plots showing the best-fit Voigt profiles and tables listing the derived column densities for each of the excited states and Doppler parameters. For the bursts where C\,{\sc i} is most prominent (GRBs\,120119A, 150403A, and 180325A) we only provide the $2\sigma$ lower limit on the total column density in Table~\ref{tab:col} since the lines are intrinsically saturated. For GRBs\,120815A, 121024A, 181020A, and 190114A we provide the derived total C\,{\sc i} column densities in Table~\ref{tab:col}. We caution, however, that since we are not able to distinguish additional narrow absorption components at the observed spectral resolution, these values should in principle only represent the lower limits on $N$(\ci) due to the possible effect of \lq hidden\rq~saturation \citep[e.g.,][]{Prochaska06}. Nevertheless, note that the inferred $b$-parameters and column densities are consistent with similar C\,{\sc i} absorption systems observed toward quasars \citep[e.g.][]{Srianand05}.

\begin{figure} 
	\centering
	\epsfig{file=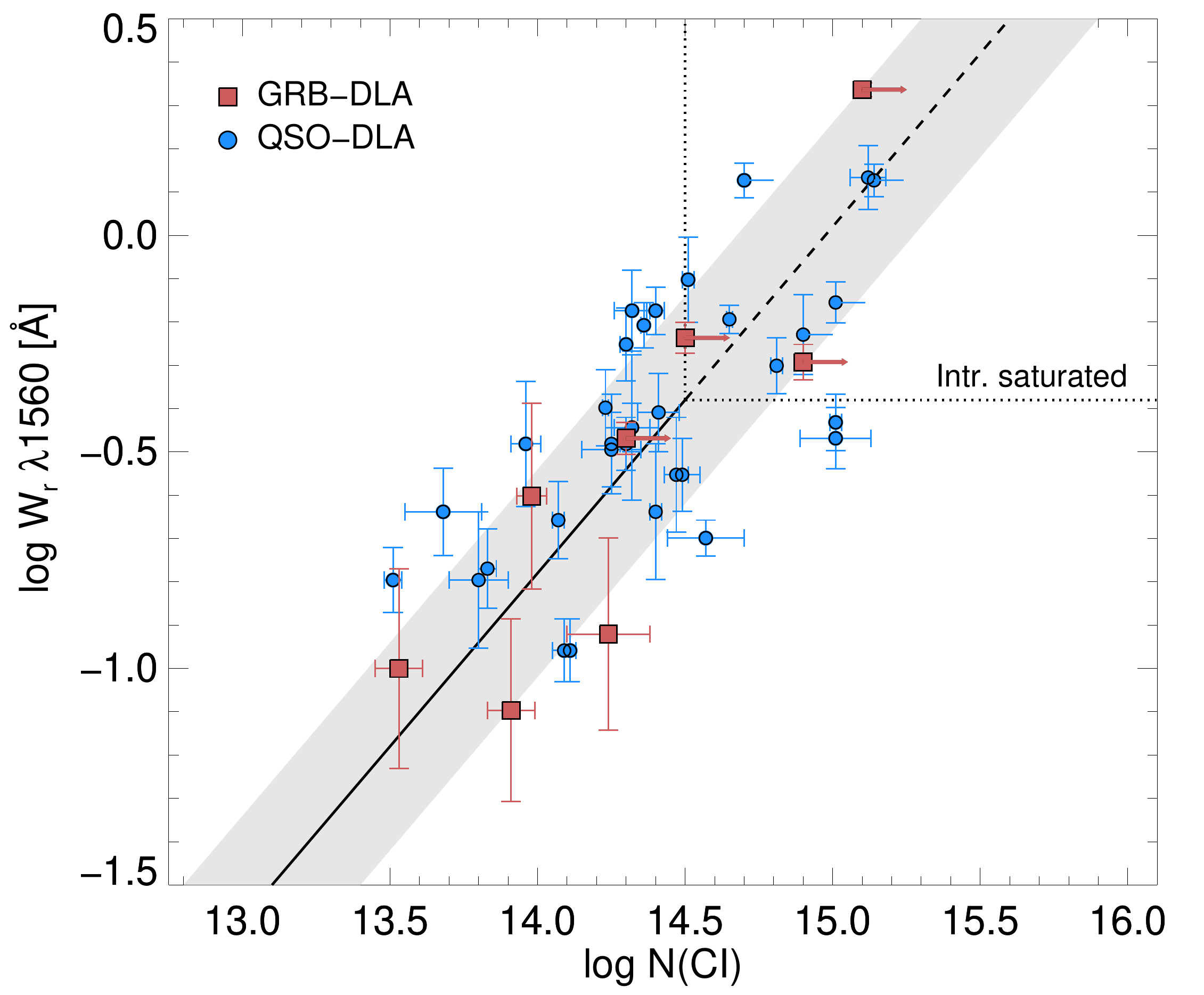,width=9cm}
	\caption{{\ci}\,$\lambda$\,1560 rest-frame equivalent width as a function of total {\ci} column density. Red squares denote GRB {\ci} absorbers from this work. Blue dots show the {\ci}-selected quasar absorbers from \cite{Ledoux15}, for which \cite{Noterdaeme18} measured the total $N$(\ci) from high-resolution spectroscopy. The black line shows the best linear fit where the $\sigma = 0.3$\,dex scatter is shown by the grey-shaded region. The approximate column density (and equivalent width) at which the intrinsic {\ci} lines saturate are marked by the dotted lines.}
	\label{fig:ciewn}
\end{figure}  

\begin{figure} 
	\centering
	\epsfig{file=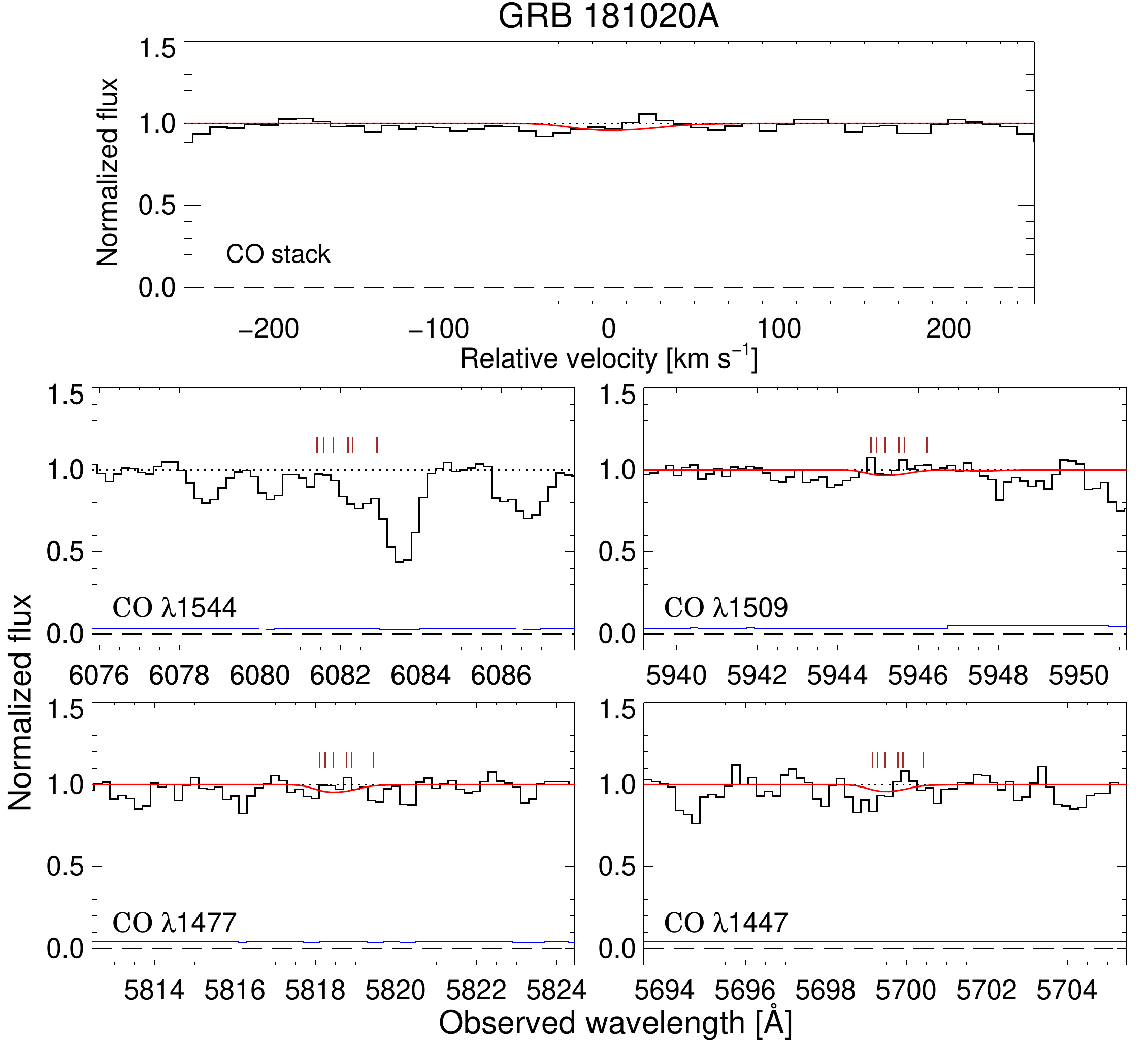,width=9cm}
	\caption{Normalized VLT/X-shooter VIS arm spectrum of GRB\,181020A showing the regions where the strongest CO band absorption lines are expected including a stacked spectrum (excluding CO\,$\lambda$\,1544). Again, the black solid lines show the data and the blue lines the associated error. Line profiles showing the derived upper limits on $N$(CO) are overplotted in red.}
	\label{fig:COspec181020A}
\end{figure}

\begin{figure} 
	\centering
	\epsfig{file=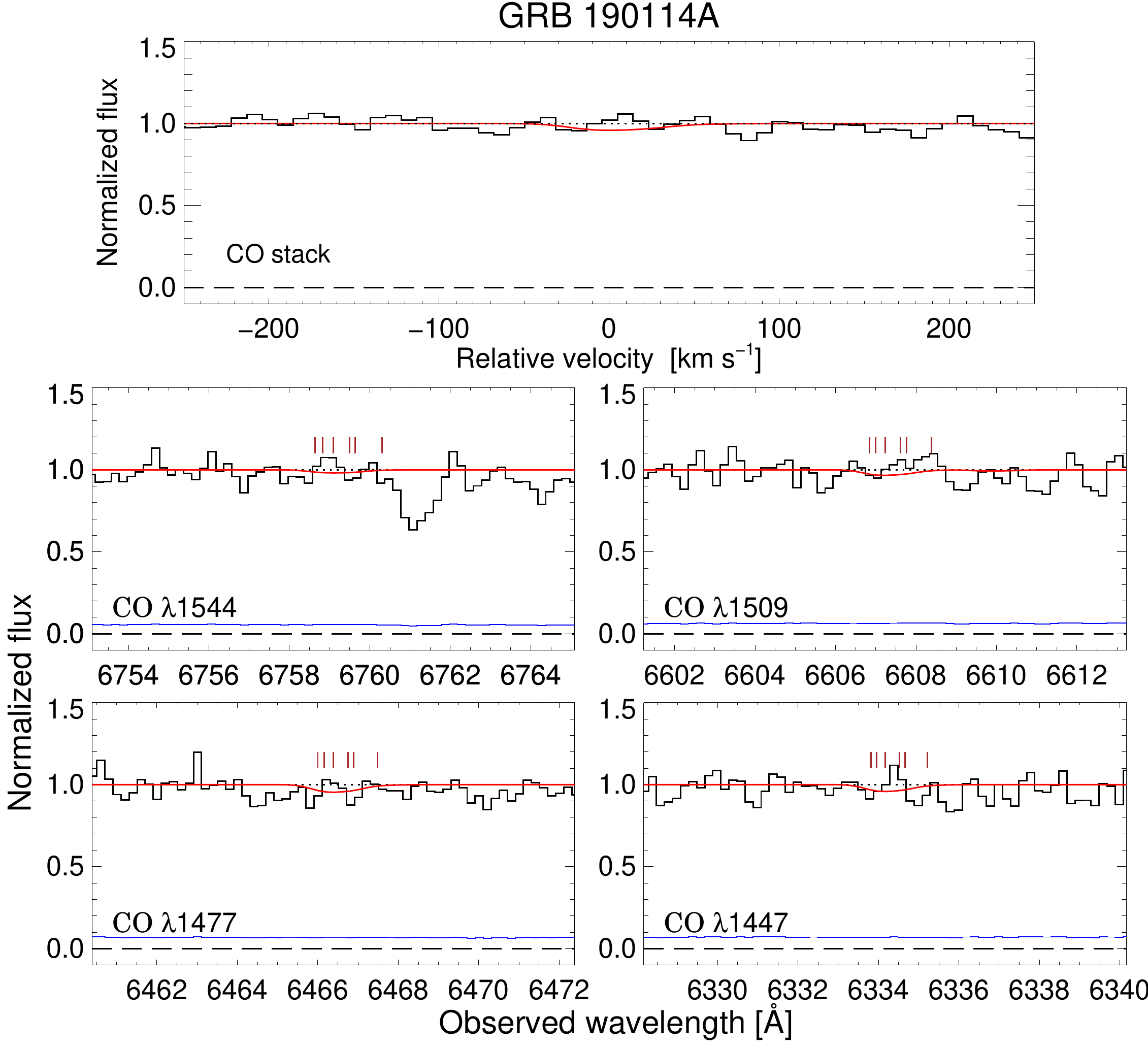,width=9cm}
	\caption{Same as Fig.~\ref{fig:COspec181020A} but for GRB\,190114A. }
	\label{fig:COspec190114A}
\end{figure}

For GRB\,080607 \citep{Prochaska09} we derive a lower limit on the total C\,{\sc i} column density based on the equivalent width measurements from \cite{Fynbo09} of $\log N$(C\,{\sc i}$)>15.1$. To estimate this abundance more precisely, we compare the C\,{\sc i} equivalent widths derived by \cite{Ledoux15} for the quasar C\,{\sc i} absorbers with the total column densities measured in the high-resolution spectra of the same absorption systems \citep{Noterdaeme18}. From a linear fit to the data (excluding the systems with $\log N$(\ci$)>14.5$, at which point the line profiles become saturated) we find a correlation of 
\begin{equation}
\log W_{\rm r}(\lambda 1560) = 0.8\log N(\textnormal{\ci}) - 11.98
\end{equation}
with a scatter of $\sigma = 0.3$\,dex (see Fig.~\ref{fig:ciewn}). This relation simply represents the evolution of equivalent width for absorption lines located on the linear part of the curve-of-growth. However, since all three fine-structure line transitions of \ci~contribute to the measured equivalent width in low-resolution spectra, this empirical relation provides a method to infer the total \ci~column density without considering the relative contributions from the three fine-structure levels. For GRB\,080607, we then estimate $\log N$(C\,{\sc i}$)=15.4\pm0.3$ based on this relation, consistent with the lower limit inferred from the equivalent width. To be conservative, we will only consider the lower limit for this GRB throughout the paper. We also note that the C\,{\sc i} equivalent widths and column densities derived for the rest of the GRB {\ci} absorbers studied here are also consistent with the correlation observed in quasar absorbers. This linear relation therefore provides a robust way of constraining the total {\ci} column density for non-saturated lines for {\ci} absorption systems observed with low-resolution spectra. 
The observed scatter of $\sigma=0.3$\,dex is likely dominated by errors, but could also reflect the varying degree of the populations in the excited fine-structure levels relative to the ground-state. 

\subsection{Carbon monoxide}

We also searched the afterglow spectra of GRBs\,181020A and 190114A for absorption features originating from carbon monoxide (CO). To date, the only detection of CO absorption lines in a GRB afterglow is towards the remarkable burst GRB\,080607 \citep{Prochaska09}. Recently, \cite{Bolmer19} derived upper limits on the CO column density for all the H$_2$-bearing GRB-DLAs examined in this study (except for GRBs\,181020A and 190114A). We also note that \citet{deUgartePostigo18} searched for CO absorption both in VLT/X-shooter and ALMA spectroscopy of GRB\,161023A but were also only able to determine upper limits (this GRB did not show features from H$_2$ down to deep limits, however). In Figs.~\ref{fig:COspec181020A} and \ref{fig:COspec190114A} we show the region of the spectra where the strongest CO band absorption lines should be located in GRBs\,181020A and 190114A, respectively, and a stack of all line complexes (excluding CO\,$\lambda$\,1544 in both cases due to blending). We do not detect any evidence of CO in either of the GRB afterglow spectra. To measure the upper limits on $N$(CO) in GRBs\,181020A and 190114A we follow the same approach as \citet{Noterdaeme18} and compute global (i.e. from the stacked spectra) $\chi^2$ values for a range of column densities, where the $3\sigma$ upper limit corresponds to the column density where the $\chi^2$ is 9. This limit is naturally more stringent than inferred locally for each band individually. For both GRBs\,181020A and 190114A we derive $3\sigma$ upper limits of $\log N$(CO$)<13.3$. The individual and stacked CO line profiles showing the upper limits on $N$(CO) are overplotted in red in Figs.~\ref{fig:COspec181020A} and \ref{fig:COspec190114A}. To be complete, we derive additional limits for GRBs\,120119A and 180325A (which were not part of the study of \citealt{Bolmer19}). A summary of the resulting upper limits on the abundance of CO for the GRBs studied in this work is provided in Table~\ref{tab:col}.

\section{Results}    \label{sec:res}

\subsection{Classification of the molecular gas-phase in GRB hosts}

The total set of H\,\textsc{i}, H$_2$, C\,\textsc{i} and CO column densities, the derived gas-phase metallicities, and visual extinctions, $A_V$, for the GRBs in our sample is provided in Table~\ref{tab:col}. All the GRB absorption systems show prominent amounts of neutral atomic hydrogen ($N$(H\,\textsc{i}$)>5\times10^{21}$\,cm$^{-2}$), comparable to the {\hi} content of extremely strong quasar DLAs \citep[ES-DLAs,][]{Noterdaeme14,Noterdaeme15a}. This further supports the hypothesis that ES-DLAs probe the neutral gas disc of intervening galaxies in quasar sightlines, similar to GRB-selected absorption systems. We observe H$_2$ column densities in the (large) range $N$(H$_2)=10^{17.2}$ to $10^{20.5}$\,cm$^{-2}$, which yield integrated molecular fractions, $f_{\mathrm{H}_2} = 2N(\mathrm{H}_2)/(2N(\mathrm{H}_2) + N(\mathrm{H\,\textsc{i}}))$, between $f_{\mathrm{H}_2} = 10^{-4.4}$ and $10^{-1.4}$. We caution that in the core of the cloud where H$_2$ (and C\,\textsc{i}) is detected, the molecular fraction is likely higher than the integrated value \citep{Balashev15} since a fraction of the atomic hydrogen is located in the \lq warm\rq~neutral medium (WNM) of the ISM.

\begin{figure} 
	\centering
	\epsfig{file=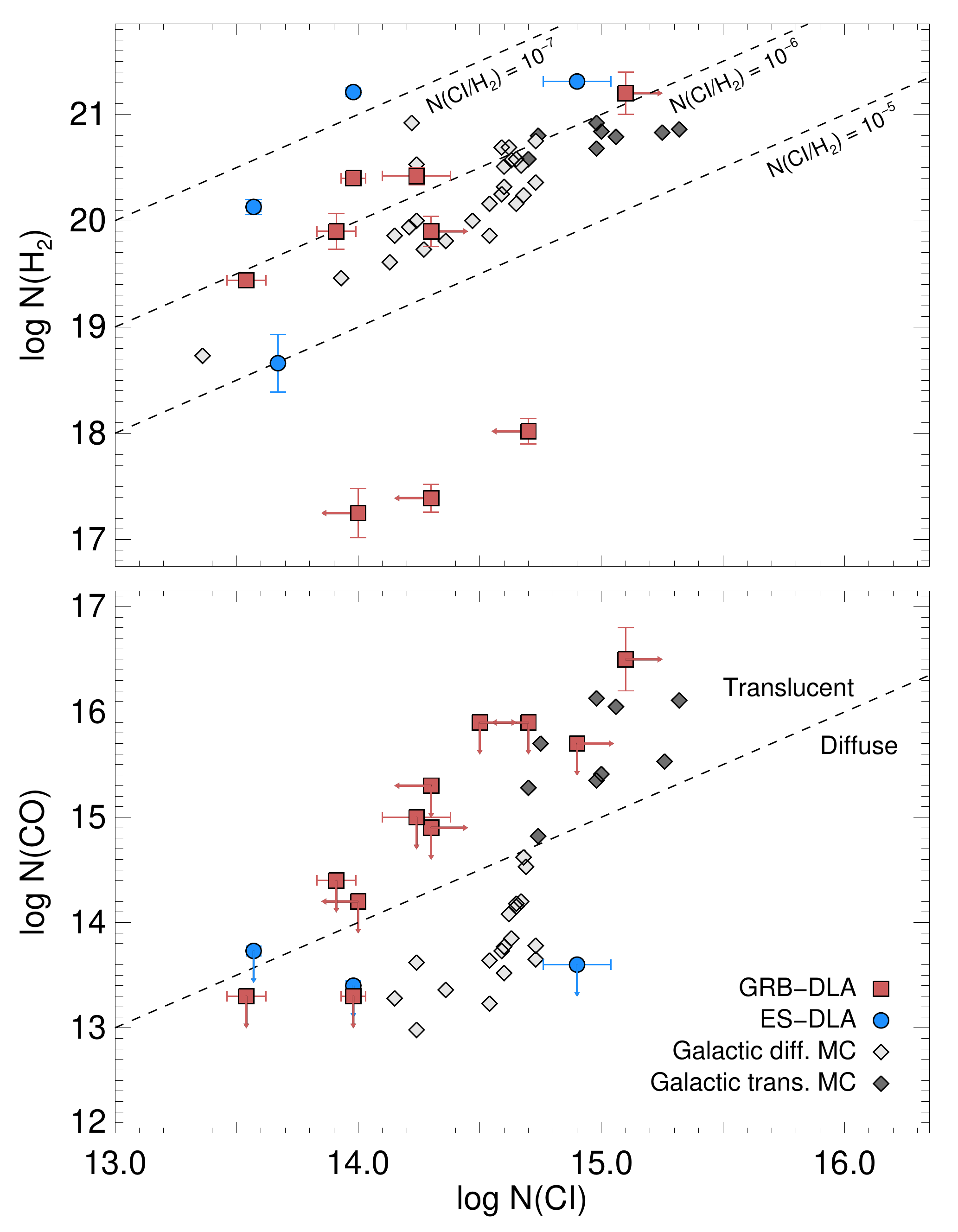,width=9cm}
	\caption{H$_2$ (top panel) and CO (bottom panel) vs C\,\textsc{i} column densities for the GRB molecular gas absorbers studied in this paper. For comparison, we also show a compiled sample of ES-DLAs with a secure detection of H$_2$ from \citet{Guimaraes12,Noterdaeme15a,Balashev17,Ranjan18}, and selected sightlines through diffuse and translucent molecular clouds in the Milky Way \citep[from][]{Burgh10}. In the top panel, a set of constant {\ci}-to-H$_2$ abundance ratios are shown for guidance. In the bottom panel, the transition region between diffuse and translucent molecular clouds at $N$(C\,\textsc{i})/$N$(CO) = 1 is shown as well.} 
	\label{fig:cih2co}
\end{figure} 

To classify the molecular gas-phase observed in GRB hosts, we follow the definition of \cite{Burgh10}. Here, diffuse molecular clouds are defined by having $N$(C\,\textsc{i}$)/N($CO$)>1$, where values below are found to trace translucent molecular gas (i.e. the transition between diffuse and dark molecular clouds, see e.g. \citealt{Snow06}). In Fig.~\ref{fig:cih2co} we plot the H$_2$ measurements and the upper limits on the CO column densities as a function of the total C\,\textsc{i} column densities for the GRB molecular gas absorbers in our sample. For comparison, we also show a small sample of ES-DLAs with a secure detection of H$_2$ \citep[compiled from][]{Guimaraes12,Noterdaeme15a,Balashev17,Ranjan18}, in addition to various diffuse and translucent molecular clouds in Galactic sightlines \citep[from][]{Burgh10}. We are not able to classify the XS-GRB molecular gas absorption systems studied here based on this classification scheme, but the typical abundance ratios of $N$(\ci$)/N$(H$_2)\approx 10^{-6}$ and the low total $N$(\ci) < 14.5 column densities are consistent with originating from diffuse molecular clouds. This is also supported by their CO/H$_2$ column density ratios and the molecular-hydrogen fractions of the systems \citep{Bolmer19}. The only exception is the host absorption system of GRB\,080607 with a relative abundance ratio of $\log N$(CO/H$_2)=-4.7\pm0.4$, consistent with originating from a translucent molecular cloud (also defined as having CO/H$_2>10^{-6}$; \citealt{Burgh10}). 

We note that in order to observe CO at a detectable level ($\log N$(CO$)\gtrsim14$), either an H$_2$ column density of $\log N$(H$_2)>20.5$ is required following the CO to H$_2$ correlation plot by \cite{Sheffer08} or a {\ci} column density of $\log N$(\ci$)>15$ (for diffuse molecular clouds), which would explain the non-detection of CO in the XS-GRB absorbers. The lower-than-solar metallicity of GRB absorbers would also result in an even lower expected CO-to-H$_2$ abundance ratio, further decreasing their detection probability.

Translucent molecular clouds can also be classified by having $A_V > 1$\,mag \citep[][see also Sect.~\ref{ssec:avmol}]{Snow06}. In our sample, only the GRBs\,120119A and 180325A (except for GRB\,080607) have dust columns consistent with this value. Unfortunately, GRB\,120119A is at too low a redshift for the Lyman-Werner bands to enter the observable UV range. For GRB\,180325A, the region of the spectrum where the potential Lyman-Werner absorption bands are present is completely suppressed by the high visual extinction \citep{Zafar18a,Bolmer19}. 

\subsection{The presence of vibrationally-excited H$_2$}

\begin{figure} 
	\centering
	\epsfig{file=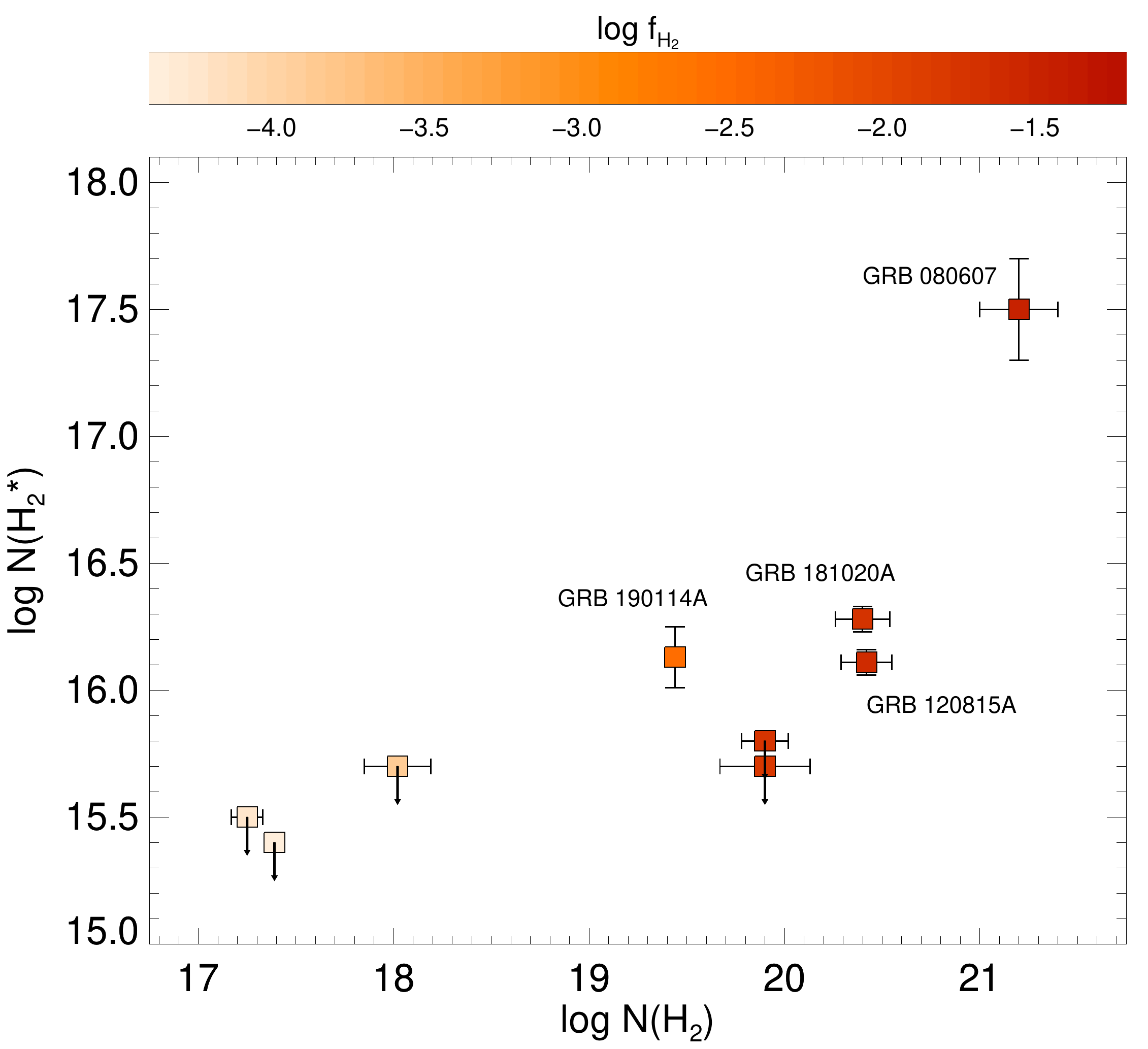,width=9cm}
	\caption{Vibrationally-excited H$_2$ (H$^*_2$) vs column densities of H$_2$ in the H$_2$-bearing GRB absorbers, color-coded as a function of the measured molecular fraction $f_{\mathrm{H}_2}$. The four GRB systems with positive detections of H$^*_2$ are marked individually.}
	\label{fig:h2vib}
\end{figure}

Similar to {\ci}, H$^*_2$ opens a potential route to establish the presence of molecular hydrogen in cases where a direct search for H$_2$ is not possible. 
Typical limitations are bursts being at too low redshifts ($z\lesssim 2$) to not cover the wavelength range bluewards of Ly$\alpha$ or significant blending of the Lyman-Werner bands with the Ly$\alpha$ forest in low-resolution spectroscopy \citep{Kruhler13,Bolmerth}. 
Fully exploiting H$^*_2$ as a molecular gas tracer, however, requires a good understanding of the observable characteristics of the H$^*_2$-bearing GRB absorbers. In Fig.~\ref{fig:h2vib} we show the positive detections and column densities of H$^*_2$ in GRBs\,080607, 120815A, 181020A and 190114A as a function of the H$_2$ column density and molecular-hydrogen fraction $f_{\mathrm{H}_2}$. 
Except for GRB\,190114A, H$^*_2$ is only detected in GRB absorbers with $\log N$(H$_2$) $\gtrsim 20$ and $f_{\mathrm{H}_2} > 0.03$. The high S/N afterglow spectrum of GRB\,190114A could explain the detection of H$^*_2$ even though the absorber has a $\sim 10$ times lower H$_2$ column density and molecular-hydrogen fraction than the other bursts with positive H$^*_2$ detections. We note though that GRBs\,121024A and 150403A, both with H$_2$ column densities and molecular fractions in the same range as GRBs~080607, 120815A, and 181020A, do not show the presence of H$^*_2$ down to $\log N$(H$^*_2$) < 15.7, which is $\sim 5$ times less abundant than in the GRB-DLAs\,120815A, 181020A, and 190114A. 

We additionally examine the dependence on the intrinsic burst luminosity for the detection probability of H$^*_2$, representing the intensity or amount of photons from the GRB producing the excitation of H$_2$. We compute the GRB energy output in the observed 15-150 keV {\it Swift}-BAT band as $E_{\rm BAT} = F_{\gamma}\,4 \pi \,d_L^2\,(1+z)^{-1}$ following \cite{Lien16}, where $F_{\gamma}$ is the observed BAT fluence in the 15–150 keV band and $d_L$ is the luminosity distance to the bursts at the given redshift. While GRBs\,080607 and 181020A are among the most luminous bursts at $z\sim 3$ with $E_{\rm BAT} > 2\times 10^{53}$\,erg \citep[see e.g.][]{Selsing19}, GRBs\,120815A and 190114A are part of the faintest {\it Swift}-detected population of bursts at their respective redshifts. 

This preliminary analysis seems to indicate that the detection probability of H$^*_2$ is likely related to several intrinsic parameters such as GRB luminosity, distance from the bursts to the absorbing molecular gas and the observational difficulty to detect H$^*_2$ with a typical low relative abundance compared to H$_2$ of $N$(H$^*_2$)/$N$(H$_2$) $\sim 10^{-4}$. This will be explored further in a follow-up paper (Bolmer et al. in preparation).

\subsection{Detecting neutral atomic carbon in GRB H$_2$ absorbers}

Molecular hydrogen observed in absorption is typically associated with C\,{\sc i} in high-redshift QSO-DLA systems \citep{Ge99,Srianand05}. C\,{\sc i}-selected quasar absorbers \citep{Ledoux15} have also been shown to always contain H$_2$ \citep{Noterdaeme18}. However, C\,{\sc i} is not ubiquitous in all H$_2$-bearing systems. The incidence rate of H$_2$ in quasar DLAs is of the order $\approx 5 - 10\%$ \citep{Ledoux03,Noterdaeme08,Jorgenson14,Balashev14,Balashev18}, whereas strong C\,{\sc i} absorption features are only found in $\approx 1\%$ of quasar absorbers \citep{Ledoux15}. In this study, we consistently find that H$_2$ is always coincident with C\,{\sc i} when the Lyman-Werner features are observable (excluding GRBs\,120119A and 180325A) in GRB-host absorbers. On the other hand, C\,{\sc i} is not detected in all the H$_2$-bearing GRB-host absorbers (as is the case for GRBs\,120327A, 120909A, and 141109A), down to similar limits as derived for the bursts with detections of \ci.

\begin{figure} 
	\centering
	\epsfig{file=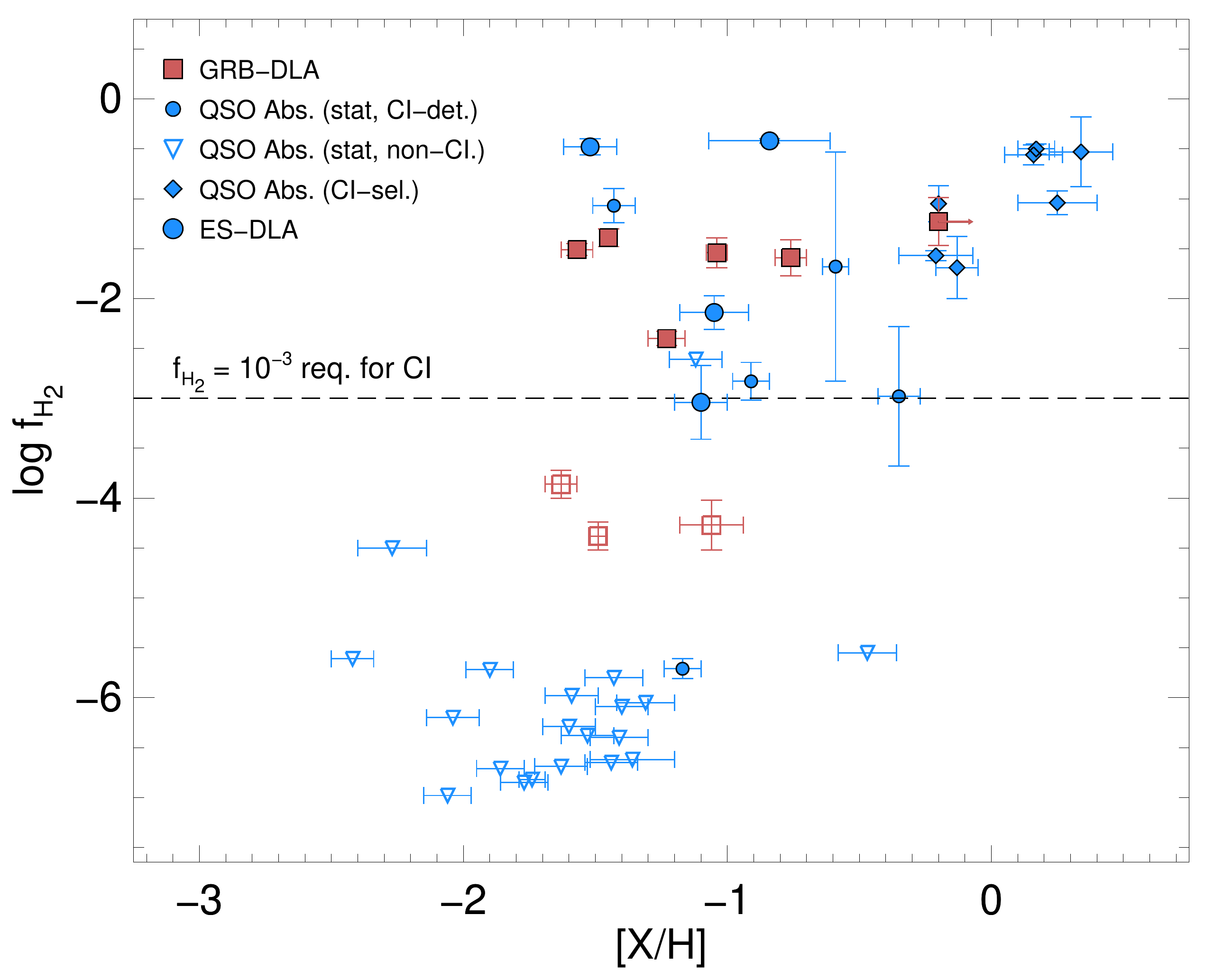,width=9cm}
	\caption{Molecular-hydrogen fraction as a function of metallicity for the H$_2$-bearing GRB absorbers (red). Filled squares denote GRBs where C\,{\sc i} is also detected, whereas empty squares represent GRBs with H$_2$ but non-detections of C\,{\sc i}. The statistical sample of quasar absorbers from \cite{Ledoux03} are shown as the small blue symbols where circles denote absorbers with H$_2$ detections and triangles show the upper limits on the molecular-hydrogen fractions of the absorbers with non-detections of H$_2$. Filled blue symbols denote quasar absorbers where \cite{Srianand05} detected C\,{\sc i} in absorption, whereas empty blue symbols represent quasar absorbers with non-detections of C\,{\sc i}. For comparison, the {\ci}-selected quasar absorbers with measurements of {\ci}, H$_2$ \citep{Noterdaeme18} and metallicities from the literature, are shown as the filled blue diamond symbols. The large blue dots represent the same ES-DLAs as shown in Fig.~\ref{fig:cih2co}.}
	\label{fig:cih2_fh2}
\end{figure}

One explanation could be that H$_2$-bearing absorbers with low metallicities consequently have less prominent amounts of carbon, below the typical detection threshold. Another possibility is that {\ci} has not been formed significantly in H$_2$ absorbers with low molecular fractions which consequently provide less shielding, such that the line-of-sight only intersects the outer-most, more diffuse regions of the cloud. To explore the conditions for C\,{\sc i} to be detected in the molecular gas-phase further, we examine the molecular-hydrogen fraction, $f_{\mathrm{H}_2}$, of the H$_2$-bearing GRB absorbers as a function of metallicity in Fig.~\ref{fig:cih2_fh2}. For comparison, we overplot the sample of quasar H$_2$ absorbers from \cite{Ledoux03} for which \cite{Srianand05} have performed a systematic search for the presence of C\,{\sc i}. We also included a sample of {\ci}-selected quasar absorbers with measurements of {\ci} and H$_2$ \citep{Noterdaeme18}, and the same sample of ES-DLAs described above. For all the GRB-host absorbers, C\,{\sc i} is only detected in systems with molecular fractions above $f_{\mathrm{H}_2} > 10^{-3}$. While C\,{\sc i} is also only observed in GRB-host absorbers with relative large metallicities ([X/H] $\gtrsim -1.5$), a similar condition appears to be required for the presence of H$_2$. The detection of {\ci} in GRB H$_2$ absorbers is, therefore, not specifically related to the metallicity of the systems. From the same samples, we also find that the total column density of C\,{\sc i} appears to be linearly correlated with the molecular-hydrogen fraction \citep[see also][]{Noterdaeme18}. 
This could indicate that a certain fraction of molecular gas is required to efficiently form and subsequently shield {\ci}. The molecular-hydrogen fraction is therefore likely the primary driver for the presence of {\ci} in H$_2$-bearing GRB-host absorbers.

\subsection{The connection between dust and molecular gas} \label{ssec:avmol}

\begin{figure} 
	\centering
	\epsfig{file=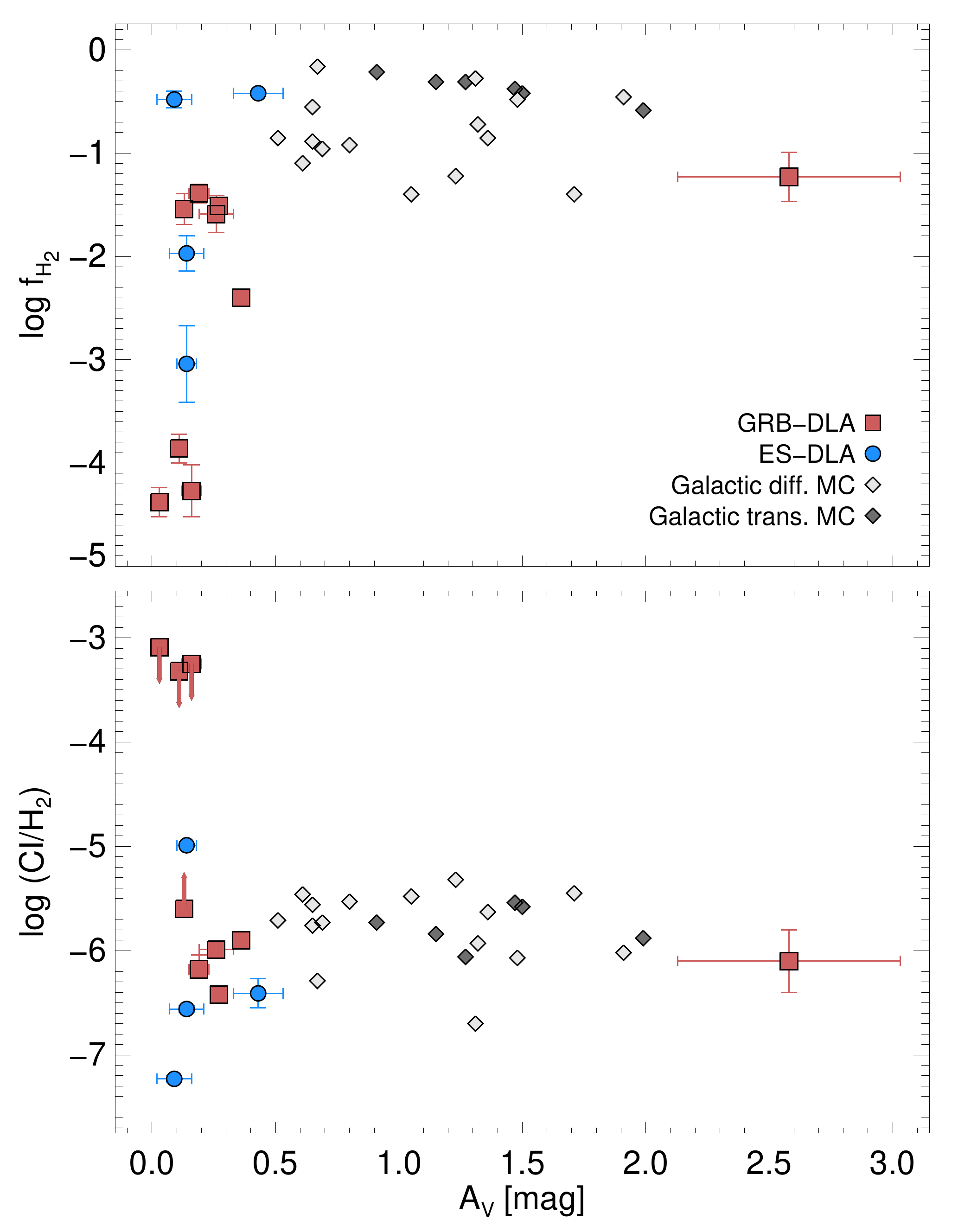,width=9cm}
	\caption{Molecular-hydrogen fraction and {\ci}/H$_2$ as a function of visual extinction, $A_V$. Red squares represent the GRBs from this work. The large blue dots represent the same ES-DLAs as shown in Fig.~\ref{fig:cih2co}. Diamond symbols show Galactic sightlines where dark gray-filled (resp. light gray-filled) symbols denote diffuse (resp. translucent) molecular clouds \citep{Burgh10}.}
	\label{fig:molav}
\end{figure}

The amount of {\ci} has been observed to be correlated with the visual extinction, $A_V$, in the line of sight to quasar and GRB absorbers \citep{Ledoux15,Ma18,Heintz19a}, suggesting a common origin for the main extinction-derived dust component and C\,{\sc i} \citep[see also][]{Heintz19b}. In Fig.~\ref{fig:molav}, we compare the measured $A_V$ of the H$_2$-bearing GRB-host absorbers to the relative molecular gas abundance ratios, in terms of $f_{\mathrm{H}_2}$ and $N$({\ci}/H$_2$). For comparison, the ES-DLAs with positive H$_2$ detections compiled from the literature are also shown, in addition to Galactic molecular-rich sightlines from the sample of \cite{Burgh10}, divided into populations of translucent or diffuse molecular clouds.

For the GRB-host absorbers, we note that there is tentative evidence for a relation between the molecular-hydrogen fraction, $f_{\mathrm{H}_2}$, and $A_V$. As an example, GRB\,080607 shows the largest dust content and highest molecular-hydrogen fraction in our sample. Computing the Kendall-rank correlation coefficient $\tau$ for $f_{\mathrm{H}_2}$ vs $A_V$ yields $\tau = 0.5$. The significance of the correlation is therefore only $1.7\sigma$. We also note that there is tentative evidence for a relation between the relative {\ci}/H$_2$ abundance ratio and $A_V$, with $\tau = 0.6$ at $1.5\sigma$ confidence, including only the GRB-host absorber with {\ci} detected in absorption. This analysis is limited by the small number of systems in our sample and the small range of (small) $A_V$ values, however, such that the derived correlations only hint at a possible connection between the amount of dust and the relative molecular gas abundance ratios. We note though, that the relatively steep rise and subsequent flattening of the relative {\ci}-to-H$_2$ abundance ratio at $A_V = 0.5$\,mag is consistent with the expected transition regime where C\,{\sc ii} is converted to {\ci} \citep{Bolatto13}.

\subsection{Kinematics}

Direct localization of the absorbing molecular gas and the explosion sites in the GRB host galaxies would provide valuable information of the immediate physical conditions of the absorbing medium. At high redshifts, however, it is challenging to obtain deep resolved images, which is required to map the varying galaxy properties accurately \citep[but see e.g.][]{McGuire16,Lyman17}. As an alternative, we can examine the relative velocity of the {\ci} and H$_2$ absorption line profiles tracing the molecular gas and compare them to the peak optical depth of the other typically observed line complexes originating from distinct gas-phase components of the ISM. Here we assume that each velocity component represents a discrete cloud in the host galaxy, located in the line of sight to the GRB. Specifically, we compare the H$_2$ and {\ci} absorption components to the line profiles from singly-ionized fine-structure transitions (typically Fe\,{\sc ii}*), and low-ionization (typically Fe\,{\sc ii}, Cr\,{\sc ii}, Mn\,{\sc ii}, or Si\,{\sc ii}) and high-ionization (N\,{\sc v}) metal lines. The relative velocity of fine-structure lines in GRB afterglow spectra carry information on the absorbing gas UV-pumped by the GRB \citep[typically at distances 0.5 -- 2 kpc from the explosion site, e.g.][]{Vreeswijk07,Vreeswijk11,DElia11}. The bulk of the metals in the neutral gas-phase of the GRB-host absorption systems is traced by the low-ionization metal lines, whereas the high-ionization lines (specifically N\,{\sc v}) have been argued to trace gas in the vicinity of the GRB \citep[within 10\,pc;][]{Prochaska08,Heintz18}.

We find that in the majority of bursts, the relative velocity of the observed H$_2$ and {\ci} line profiles are kinematically \lq cold\rq, being offset by $\delta v \lesssim 20$\,km\,s$^{-1}$ from the strongest low- and high-ionization and fine-structure line components. Due to the medium resolution of the data, these low offsets should be treated as being consistent with zero. By association, we argue that the different gas-phase components probed by the various lines suggest that they originate from the same approximate region as the bulk of the neutral gas. The fact that N\,{\sc v} is coincident with the bulk of the neutral absorbing gas in these GRB-host absorption systems is, however, likely due to their high $N$(H\,{\sc i}), which might confine the gas to the central regions of the host galaxy \citep{Heintz18} and is likely related to the GRB event itself. 

The two exceptions are GRBs\,121024A and 150403A. \cite{Friis15} showed that for GRB\,121024A, the redshift of the Lyman-Werner lines of molecular hydrogen is coincident with the strongest low-ionization metal components, which we also confirm from the {\ci} line profiles. However, the absorption profiles reveal an additional metal line complex at $\delta v \approx -400$\,km\,s$^{-1}$ which in turn is coincident with the fine-structure line transitions from Fe\,{\sc ii}* and Ni\,{\sc ii}*. The absorbing molecular gas in the GRB host is therefore likely at even greater distances from the explosion site than the gas photoexcited by the GRB \citep[which was found to be at a distance of $\approx 600$\,pc;][]{Friis15}. For GRB\,150403A, we find that the peak optical depth of {\ci} is coincident with the strongest components from the low-ionization and fine-structure lines, but offset by $\delta v \approx -30$\,km\,s$^{-1}$ from N\,{\sc v} \citep[see also][]{Heintz18}.

\section{Discussion} \label{sec:disc}

We have now established that GRB-host absorbers can be used to probe the diffuse molecular gas-phase in their high-$z$ star-forming galaxies. Additionally, we were able to quantify the defining characteristics of the subset of H$_2$-rich absorbers showing the presence of {\ci} and H$^*_2$. This section is aimed at further understanding the physical properties of the molecular gas in high-$z$ GRB absorbers and explore the possible consequence of a severe dust bias in the detection of H$_2$-bearing GRB absorbers. 

\subsection{Excitation temperature} \label{ssec:tex}

One of the key physical properties of the cold neutral medium (CNM) is the temperature, which is typically found to be in the range 30 -- 100\,K for diffuse molecular clouds \citep{Snow06}. For the H$_2$-bearing GRB-host absorbers we can infer the excitation temperature of the molecular gas from
\begin{equation}
\frac{N(\mathrm{H}_2,\,j)}{N(\mathrm{H}_2,\,i)} = \frac{g(\mathrm{H}_2,\,j)}{g(\mathrm{H}_2,\,i)}\exp^{-E_{ij}/kT_{ij}}~,
\end{equation}
where $g$ is the spin statistical weight $g(j) = 2j+1$, $E_{ij}$ is the energy difference between levels $i$ and $j$, and $T_{ij}$ is the excitation temperature. The temperature determined from the lowest two rotational levels ($J=0$ and $J=1$), $T_{01}$, is found to be a good representation of the overall kinetic temperature of the thermalized molecular gas \citep{Roy06}, whereas higher rotational levels typically indicate larger excitation temperatures due to molecule formation and/or UV pumping. 

\begin{figure} 
	\centering
	\epsfig{file=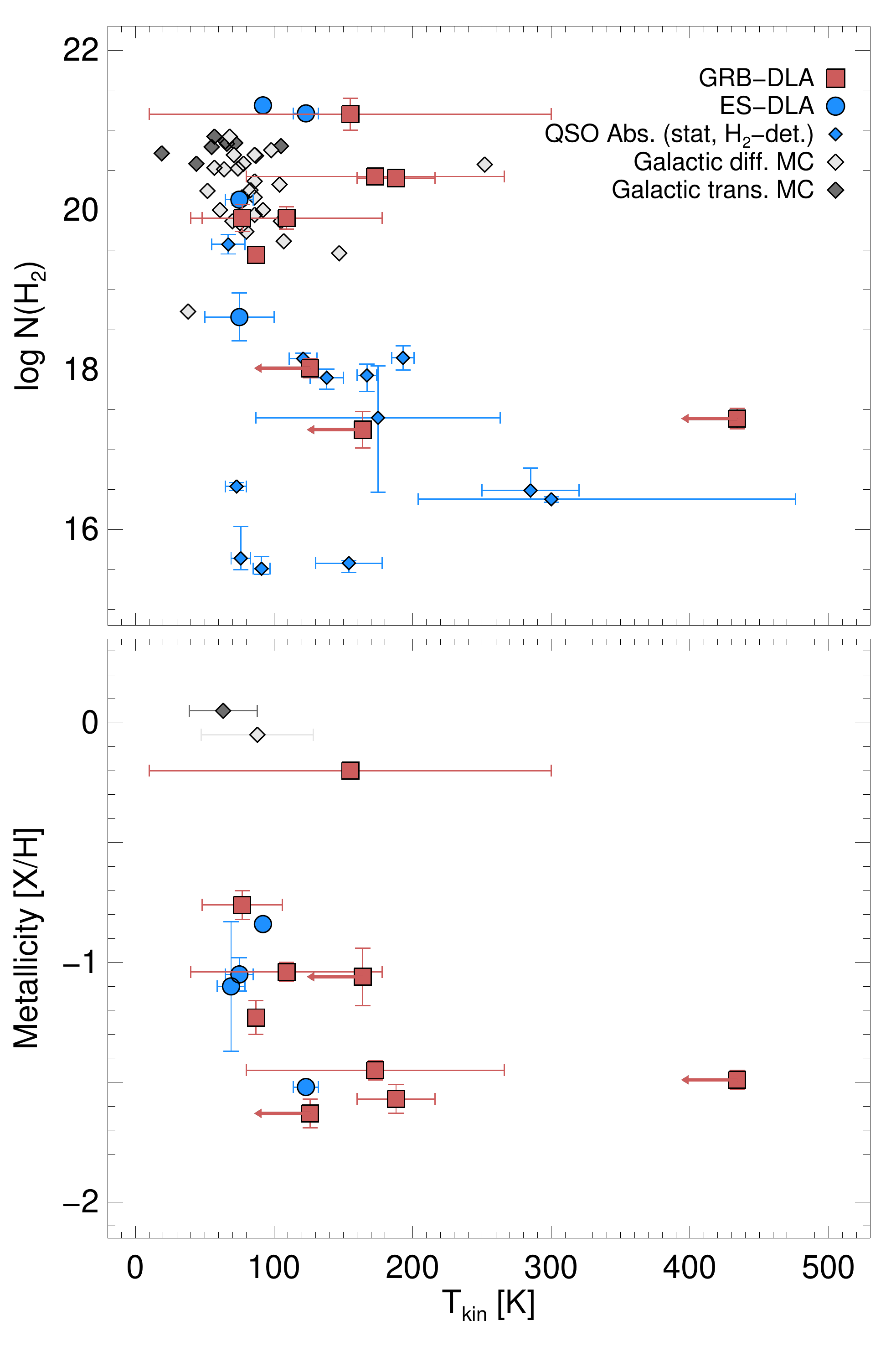,width=9cm}
	\caption{Measurements of the kinetic temperature from the H$_2$ rotational excited states as a function of total H$_2$ column density (top) and metallicity (bottom). Red squares represent the GRBs from this work, where the arrows mark the upper limits on $T_{\rm kin}$ derived from $T_{12}$ (see text). The small blue diamond symbols denote quasar DLAs from the sample studied by \cite{Srianand05}, where large blue dots represent the same ES-DLAs shown in Fig.~\ref{fig:cih2co}. The grey-shaded diamond symbols in the top panel again show selected sightlines through diffuse and translucent molecular clouds in the Milky Way \citep[from][]{Burgh10}. In the bottom panel, only the strongest DLAs with $N$(\hi) > $10^{21.7}$\,cm$^{-2}$ are shown. The grey-shaded diamond symbols here represent the mean $T_{\rm kin}$ for Galactic diffuse and translucent molecular clouds (with error bars denoting the standard deviation), arbitrarily placed at solar metallicities.} 
	\label{fig:temp}
\end{figure} 

For the GRBs\,120815A, 121024A, 150403A, 181020A, and 190114A, where the H$_2$ column densities in the two lowest rotational states are well constrained, we are able to robustly measure $T_{01}=T_{\rm kin}$. For the other systems with H$_2$ detections (i.e. GRBs\,120327A, 120909A, and 141109A), the $J=0$ state is not well constrained by the fit. For these bursts, the derived $T_{01}$ becomes negative, which could suggest that the assumption of equilibrium may not be correct in these cases. To overcome this, we instead compute the excitation temperature from the first two excited states, $T_{12}$, but only consider those as upper limits for the kinetic temperature since it is typically found that $T_{12} > T_{01}$ \citep[e.g.][]{Srianand05}. For GRB\,080607, \cite{Prochaska09} estimated an excitation temperature in the range $T_{\mathrm{ex}} = 10 - 300$\,K. 
The inferred molecular gas temperatures and upper limits are shown in Fig.~\ref{fig:temp} as a function of $N$(H$_2$). Here, we also compare the GRB H$_2$ absorbers to the H$_2$-bearing quasar absorbers examined by \cite{Srianand05}, the sample of H$_2$-bearing ES-DLAs and Galactic molecular clouds \citep{Burgh10}. In general, the H$_2$-bearing GRB absorbers contain the largest H$_2$ column densities observed at high redshift, comparable to quasar ES-DLAs and Galactic molecular clouds. We infer kinetic temperatures in the range $T_{\mathrm{kin}} \approx 100 - 300$\,K, consistent with the majority of H$_2$-bearing quasar absorbers \citep[][]{Srianand05,Balashev17}. We note that there appears to be tentative evidence for the highest metallicity GRB-host and ES-DLA systems to show lower excitation temperatures at a given H$_2$ column density (see Fig.~\ref{fig:temp}), but only at low significance due to the limited data at hand. Computing the Spearman $\rho$ and Kendall-rank $\tau$ correlation coefficients for $T_{\rm ex}$ vs [X/H] yield $\rho = -0.67$ and $\tau = -0.49$, such that the correlation significance is $2.01\sigma$ (considering only the measurements and excluding limits). 

We also examine the higher rotational transitions of H$_2$ in the afterglow spectra of GRBs\,181020A and 190114A, which might provide clues on the more external layers of the cloud and the incident UV flux. The column densities of the $J\ge 4$ transitions are not well-constrained in either case, so instead of fitting the individual line transitions we produce a synthetic spectrum, including all $J$ transitions up to $J=7$, to match to the data. For both GRBs, we fix the redshift and total H$_2$ column density to the already-determined values and only increase the excitation temperature. We assume a conservative $b$-parameter of $b=10$\kms~in both models, since the high-$J$ levels are typically found to show broader features than the low-$J$ transitions \citep[see, e.g.,][]{Noterdaeme07}. 
For GRB\,181020A, the spectrum does not show any clear indication of features arising from the $J\ge 4$ transitions, suggesting that none of the high-$J$ transitions are significantly populated. Based on our model, we estimate that a maximum high-$J$ excitation temperature of $T_{\rm ex} \sim 300$\,K is consistent with the observed spectrum. For GRB\,190114A, we find that the spectrum is consistent with high-$J$ features arising from a warmer medium, constrained to $T_{\rm ex} \sim 500$\,K. This would indicate that the intensity of the ambient UV field in the host of GRB\,190114A is higher compared to the host of GRB\,181020A. It is in principle possible to indirectly measure the ambient UV flux from the fine-structure transition of C\,{\sc ii}*\,$\lambda\,1335$ \citep{Wolfe03}. However, in both the afterglow spectra of GRBs\,181020A and 190114A this feature is either saturated or blended with C\,{\sc ii}\,$\lambda$\,1334.

\subsection{The implications of a dust bias for the detection of H$_2$} \label{ssec:h2bias}

With the increased number of known H$_2$-bearing GRB absorbers, it is clear that the first apparent bias against this subpopulation \citep{Tumlinson07,Whalen08,Ledoux09} is partly alleviated. This is largely owing to the more sensitive, higher-resolution X-shooter spectrograph, with which a large statistical sample of GRBs has been obtained \citep{Selsing19}. Expanding the discussion from \citet{Kruhler13}, we now wish to quantify to what extent the XS-GRB sample is biased against the most metal- and dust-rich H$_2$-bearing GRB-host absorption systems. Specifically whether a significant dust bias exists decreasing the H$_2$ detection probability in these systems.

\begin{figure} 
	\centering
	\epsfig{file=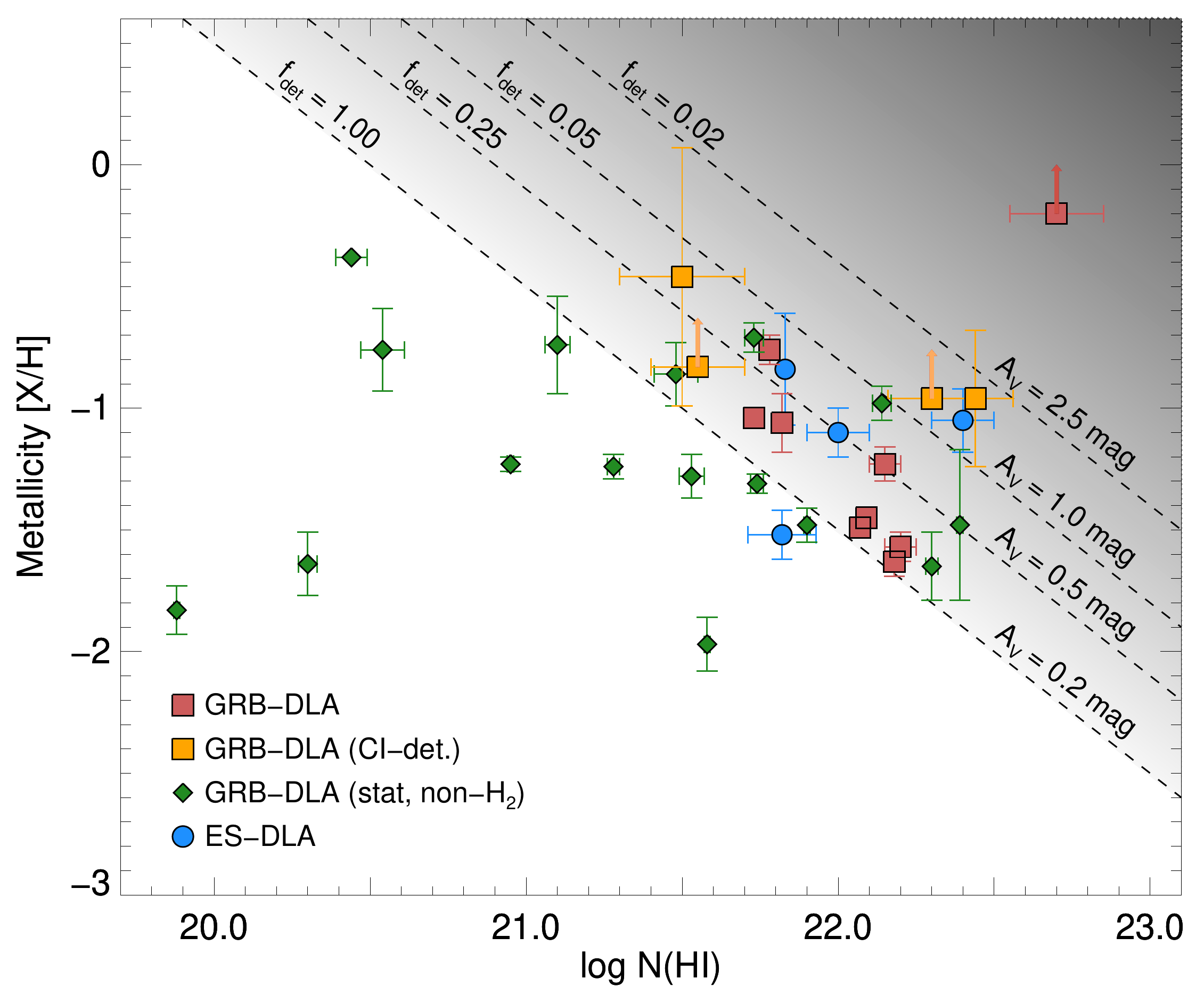,width=9cm}
	\caption{Gas-phase metallicity as a function of {\hi} column density. Filled red squares represent the GRBs from this work with positive H$_2$ detections, orange squares denote {\ci}-bearing GRBs (where $\mathrm{H}_2$ could not be constrained) from this work or \citet{Heintz19a} and the green diamond symbols show the statistical GRB-DLA sample from \citet{Bolmer19} without H$_2$ detections. For comparison, the large blue dots show the same ES-DLAs as shown in Fig.~\ref{fig:cih2co}. The gray-shaded region and overplotted dashed lines show the expected increase in $A_V$ for a given dust-to-metals ratio, in addition to the estimated detection probability $f_{\rm det}$ of dusty $\mathrm{H}_2$-bearing GRB absorbers (see Sect.~\ref{ssec:h2bias} for further details).}
	\label{fig:h2bias}
\end{figure}

To do so, we compare the $A_V$ distribution of the statistical sample of XS-GRBs, from which \citet{Bolmer19} searched for H$_2$, to an unbiased sample of GRB afterglows \citep{Covino13}. We normalize the two distributions to the number of bursts with $A_V < 0.2$\,mag (which we expect the XS-GRB sample to at least be complete to) and compute the fraction of H$_2$-bearing GRBs to the number of bursts in the unbiased sample in bins of $A_V = 0.2 - 0.5$, $0.5 - 1.0$, and $1.0 - 2.5$\,mag. We find that already at $A_V = 0.2 - 0.5$\,mag, the detection probability of the bursts in the XS-GRB sample only constitutes $\sim 25\%$ of the underlying distribution. At $A_V = 0.5 - 1.0$\,mag we estimate the fraction of uncovered GRB H$_2$ absorbers to be $\sim 5\%$, based only on the detection of {\ci} in GRB\,120119A which was not part of the statistical H$_2$ sample so this fraction effectively only serves as an upper limit. Similarly for the $A_V = 1.0 - 2.5$\,mag range, we estimate the detection probability to be 2\%, based on the single detection of H$_2$ in GRB\,080607 in the unbiased sample of $\sim 50$ bursts by \citet{Covino13}. Again, the detection probability is likely lower since none of the H$_2$-bearing XS-GRBs show $A_V$ in this range. This is illustrated in Fig.~\ref{fig:h2bias}, where we show the metallicity as a function of {\hi} column density of the GRB absorbers and compare to the expected dust extinction for a given dust-to-metals ratio \citep{Zafar13}. These estimates do not take into account the increased difficulty of detecting H$_2$ in faint bursts (either intrinsically or due to overall stronger dust absorption) and also do not include the possibility of steep extinction curves. For example, GRB\,140506A \citep{Fynbo14,Heintz17} would be practically invisible at optical wavelengths if it were located at $z \gtrsim 2$. The dust bias in the observed XS-GRB sample might therefore be even more severe than the simple estimates provided here.

We wish to emphasize though, that if a spectrum of a burst similar to GRB\,180325A (with strong {\ci} absorption and $A_V \sim 1.5$\,mag) was obtained with higher S/N spectroscopy, the H$_2$ features might have been revealed as well. Conversely, this further demonstrates the versatility of using {\ci} as an alternative tracer of molecular gas, even in very dust-reddened sightlines. We conclude that the large majority of dusty ($A_V > 0.2$\,mag) H$_2$-bearing GRBs are missed due to a significant dust bias. This confirms the proposal by \citet{Ledoux09}, that GRB-host absorber samples are likely to be biased against dusty and metal-rich sightlines. Only in the cases of rare, extremely luminous afterglows \citep[such as GRB\,080607;][]{Prochaska09,Perley11} is it possible to detect H$_2$ in the most dust-obscured afterglows. 

\section{Conclusions} \label{sec:conc}

We have presented optical to near-infrared VLT/X-shooter spectra of the afterglows of GRBs\,181020A and 190114A at $z=2.938$ and $z=3.376$, respectively. Both sightlines are characterized by strong DLAs and substantial amounts of molecular hydrogen with $\log N$(\hi, H$_2$) = $22.20\pm 0.05,~20.40\pm 0.04$ (GRB\,181020A) and $\log N$(\hi, H$_2$) = $22.15\pm 0.05,~19.44\pm 0.04$ (GRB\,190114A). Both GRB-host absorption systems show relatively high molecular fractions of $f_{\rm H_2} = 0.4 - 3\%$, characteristic of Galactic diffuse molecular gas and consistent with the majority of H$_2$-bearing quasar absorbers at high-$z$. These two cases represent only the eighth and ninth unambiguous detection of H$_2$ in GRB-host absorption systems. We measure gas-phase metallicities of [Zn/H] = $-1.57\pm 0.06$ and $-1.23\pm 0.07$, relative depletion abundances of [Zn/Fe] = $0.67\pm 0.03$ and $1.06\pm 0.08$, and visual extinctions of $A_V = 0.27\pm 0.02$ mag and $0.36\pm 0.02$ mag, for GRB\,181020A and GRB\,190114A, respectively. While the metallicities of the two systems are relatively low and comparable to typical GRB-host absorbers, their metal column densities, $\log N$(\hi) + [Zn/H], are among the highest in the general GRB-host absorber population. They are also both well above the apparent GRB H$_2$ detection threshold of $\log N$(\hi) + [Zn/H] $> 20.5$ \citep{Bolmer19}.

In addition to molecular hydrogen, we also detect absorption features from neutral atomic carbon and vibrationally-excited H$_2$ in both afterglow spectra of GRBs\,181020A and 190114A. To complement the analysis of these alternative molecular gas tracers, and to explore the conditions for these rarer absorption features to arise, we systematically searched all the H$_2$-bearing GRB absorbers from \citet{Bolmer19} for the presence of {\ci} or H$^*_2$ and measured or provided limits on the respective column densities. We found that {\ci} and H$^*_2$ are efficient tracers of H$_2$-rich GRB-host absorbers, but also that H$_2$ does not guarantee the presence of either. 
First, we explored the conditions required to detect {\ci} in the H$_2$-bearing GRB absorbers and we found that an apparent threshold of the overall molecular-hydrogen fraction of $f_{\rm H_2} > 10^{-3}$ is essential. The total {\ci} column density is also found to be linearly connected with $f_{\rm H_2}$. The defining characteristic for the presence of H$^*_2$ is less clear, likely because it depends on several parameters such as the H$_2$ abundance, GRB luminosity and distance to the absorbing molecular gas. This somewhat limits the applications of {\ci} and H$^*_2$ as overall efficient molecular gas tracers. On the other hand, identifying absorption features from {\ci} or H$^*_2$ provides indirect evidence of large H$_2$ abundances, even in the absence of the Lyman-Werner H$_2$ features (e.g. due to low redshifts, large dust content or low spectral resolution). We also compared the kinematics of the absorption lines from the molecular gas tracers {\ci} and H$_2$ to the low- and high-ionization and fine-structure absorption features typically observed in GRB-host absorbers. We found that {\ci} and H$_2$ are in most cases kinematically \lq cold\rq, thus likely confined to the same proximate region as the bulk of the metals producing the strongest low- and high-ionization absorption features.

Based on the now nine positive detections of H$_2$ in GRB-host absorbers, we examined the typical excitation temperatures of the molecular gas, constrained from the two lowest rotational levels of H$_2$ ($J=0,1$). For the systems in our sample we inferred temperatures in the range $T_{\rm ex} = 100 - 300$\,K. A more careful analysis of the high-$J$ H$_2$ transitions in GRBs\,181020A and 190114A revealed tentative evidence of a slightly warmer component with up to $T_{\rm ex} = 300$ and 500\,K, respectively. Finally, we determined the probability of detecting H$_2$ in the XS-GRB afterglow sample \citep{Selsing19} as a function of $A_V$. Even in moderately extinguished sightlines with $A_V \gtrsim 0.2$\,mag, the number density of GRB H$_2$ absorbers drops to $\sim 25\%$ compared to an unbiased sample of GRB afterglows. This suggests that while the XS-GRB afterglow survey has been successful in recovering a significant number of H$_2$-bearing GRB absorbers \citep{Bolmer19}, the most dust-obscured systems are still missed due to a non-negligible dust bias.

In summary, GRB-host absorbers provide detailed information about the characteristics and physical properties of the diffuse molecular gas-phase in the ISM of star-forming galaxies during the peak of cosmic star-formation. While the absorption features of the molecular gas tracers are typically only detected at UV/optical wavelengths, there are promising prospects of detecting them at sub-mm wavelengths in ALMA spectroscopy as well \citep{deUgartePostigo18}. Connecting the properties inferred from absorption-line analyses to the CO line emission at sub-mm wavelengths would also provide unparalleled insight into the conditions and physical processes fuelling star-formation at high redshift. So far, only a small number of high-$z$ GRB host galaxies have been detected in emission from CO \citep{Michalowski18,Arabsalmani18}, though without any constraints on the molecular gas properties from absorption. Targeting the CO emission lines of this sample of H$_2$-bearing GRB-host galaxy absorbers would provide a natural unification of the two approaches. In the near future, identification of the vibrational and ro-vibrational H$_2$ emission lines will also be possible with the {\it James Webb Space Telescope} \citep{Kalirai} at $z\gtrsim 2$ where H$_2$ can be detected in absorption \citep{Guillard15}, which so far has only been detected in a single, $z\sim 0.1$ GRB host \citep{Wiersema18}. This combined analysis of molecular gas in line-of-sight GRB afterglow spectra and integrated host galaxy spectra will also greatly benefit the typically more extensive emission-selected CO galaxy surveys, and significantly improve our understanding of the connection between cold and molecular gas observed in absorption and emission.


\begin{acknowledgements}
We would like to thank the referee for a clear, concise and timely report.
KEH and PJ acknowledge support by a Project Grant (162948--051) from The Icelandic Research Fund. PN and JKK acknowledge support from the French {\sl Agence Nationale de la Recherche} under contract  ANR-17-CE31-0011-01 (Projet "HIH2", PI Noterdaeme) and are grateful to the European Southern Observatory for hospitality and support during a visit to the ESO headquarters in Chile. The Cosmic Dawn Center is funded by the DNRF. AdUP, CCT, DAK and LI acknowledge support from the Spanish research project AYA2017-89384-P, and from the State Agency for Research of the Spanish MCIU through the "Center of Excellence Severo Ochoa" award for the Insituto de Astrof\'isica de Andaluc\'ia (SEV-2017-0709). AdUP and CCT acknowledge support from Ram\'on y Cajal fellowships (RyC-2012-09975 and RyC-2012-09984). LI acknowledges support from a Juan de la Cierva Incorporaci\'on fellowship (IJCI-2016-30940).
\end{acknowledgements}

\bibliographystyle{aa}
\bibliography{ref}

\appendix

\section{Gas-phase abundances and dust extinction toward GRBs\,181020A and 190114A}

\begin{figure} 
	\centering
	\epsfig{file=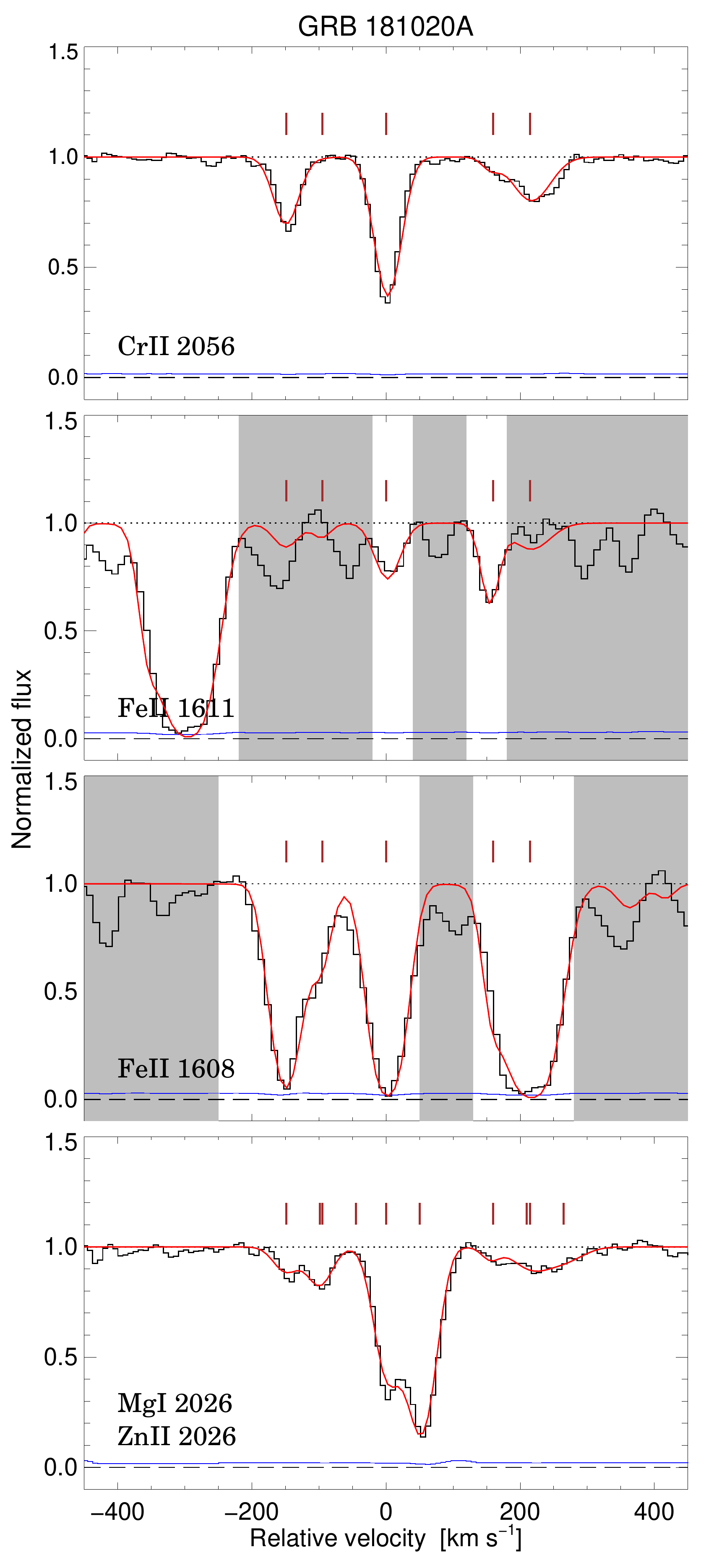,width=8.5cm}
	\caption{Normalized VLT/X-shooter spectrum of GRB\,181020A in velocity space, centred on the strongest component at $z=2.9379$. The black solid line shows the spectrum and the associated error is shown in blue. The best-fit Voigt profiles are indicated by the red solid lines. The identified velocity components are marked above each of the absorption profiles. Gray shaded regions were ignored in the fit. These lines are representative of the typical low-ionization metal lines in GRB\,181020A, showing most clearly the overall velocity structure.}
	\label{afig:181020a_met}
\end{figure}

\begin{figure} 
	\centering
	\epsfig{file=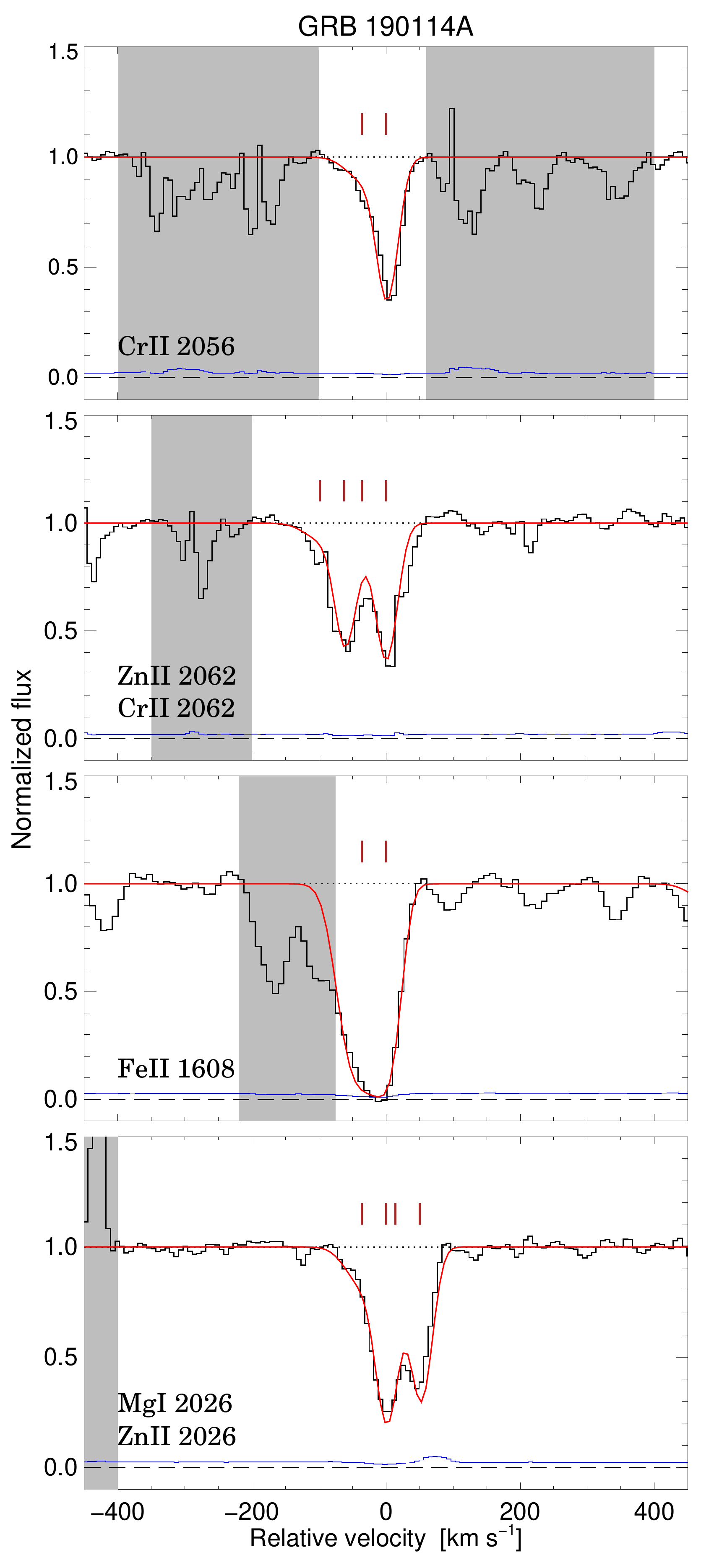,width=8.5cm}
	\caption{Same as Fig.~\ref{afig:181020a_met} but for GRB\,190114A, centred on $z=3.3764$.}
	\label{afig:190114a_met}
\end{figure}

\begin{figure} 
	\centering
	\epsfig{file=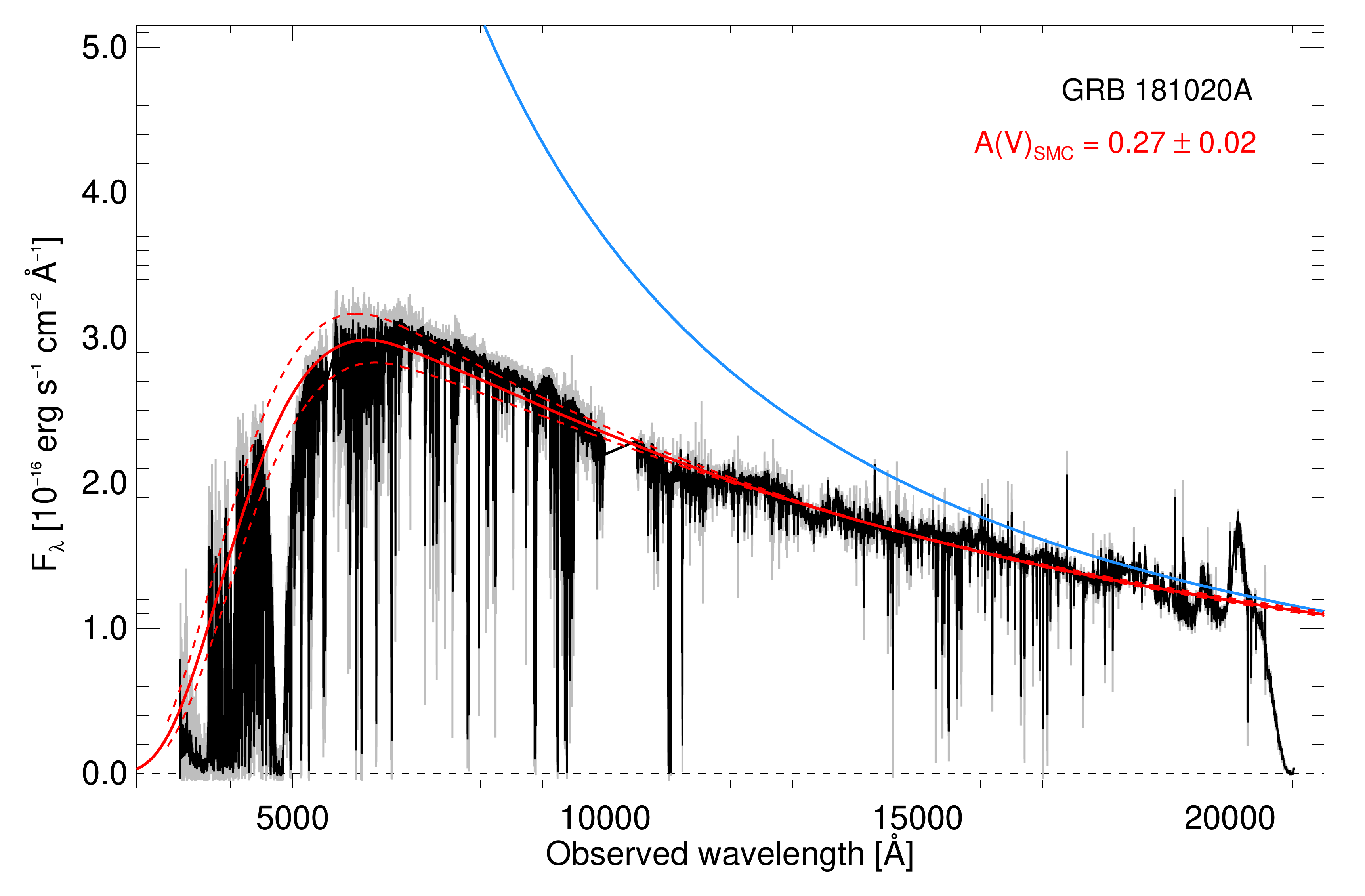,width=9cm}
	\caption{Extinction curve fit for GRB\,181020A. The raw combined VLT/X-shooter spectrum is shown in grey, overplotted with a binned version to enhance the continuum trace. The best-fit SMC-like extinction with $A_V = 0.27\pm 0.02$\,mag is shown by the red solid line, where the error on the fit is shown by the red dashed lines. The best-fit intrinsic spectral slope is overplotted as the solid blue line.}
	\label{afig:181020a_ext}
\end{figure}

\begin{figure} 
	\centering
	\epsfig{file=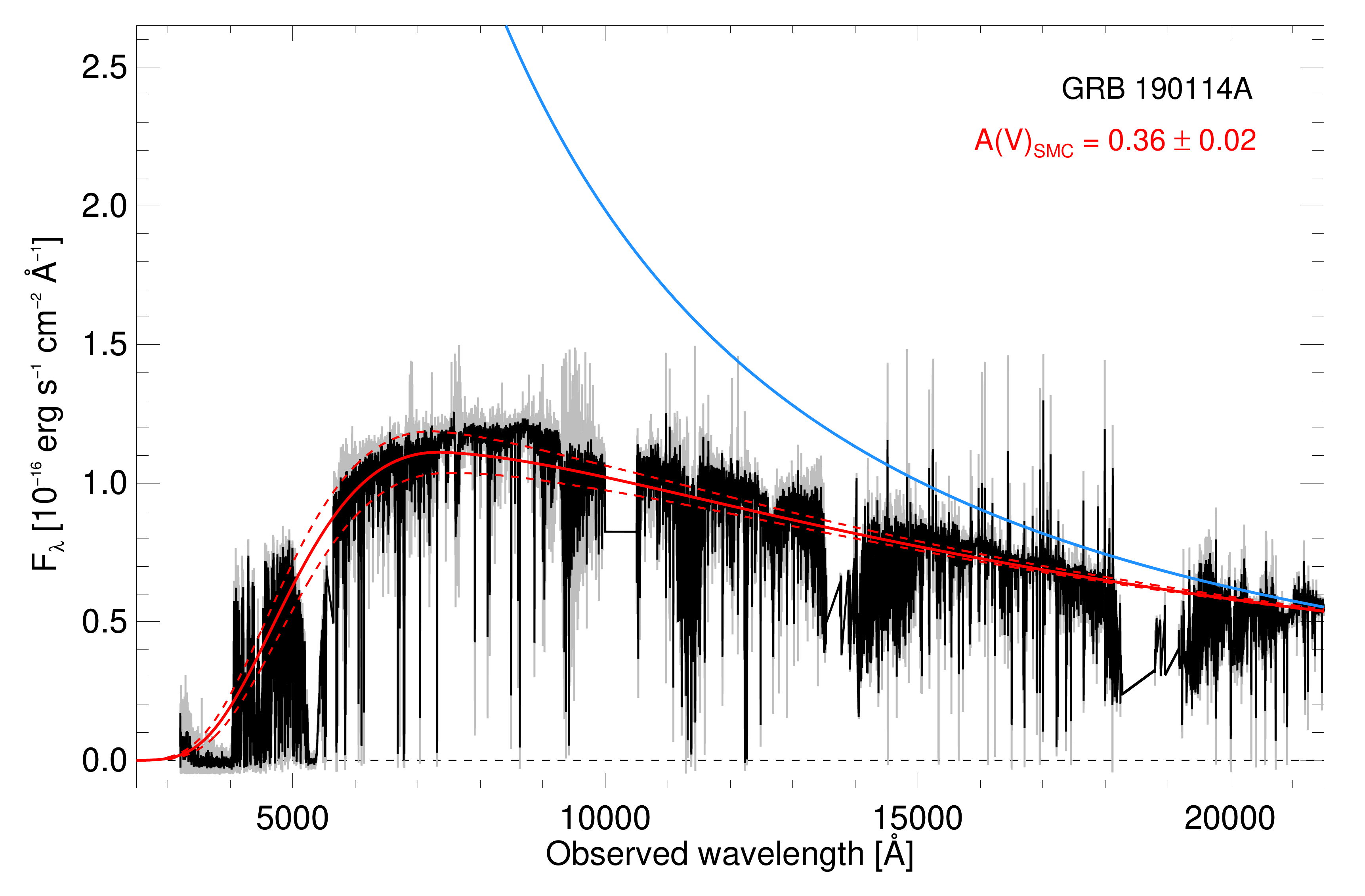,width=9cm}
	\caption{Same as Fig.~\ref{afig:181020a_ext} but for GRB\,190114A, with a best-fit SMC-like extinction of $A_V = 0.36\pm 0.02$\,mag.}
	\label{afig:190114a_ext}
\end{figure}

Here a subset of the low-ionization metal lines observed in the afterglow spectra of GRB\,181020A (Fig.~\ref{afig:181020a_met}) and GRB\,190114A (Fig.~\ref{afig:190114a_met}) is shown. We fitted several other transitions, including single-ionized elements not shown in these plots to constrain the velocity components of the absorption line profiles. However, we here only show few selected lines that best represent the overall velocity structure of the line profiles and focus on the elements used to determine the gas-phase abundance and depletion (e.g. Zn\,{\sc ii} and Fe\,{\sc ii}). In Figs.~\ref{afig:181020a_ext} and \ref{afig:190114a_ext} we show the best-fit extinction observed toward GRBs\,181020A and 190114A, respectively. Both sightlines can be modelled by a smooth, SMC-like extinction curve and are moderately reddened with $A_V = 0.27\pm 0.02$ mag and $A_V = 0.36\pm 0.02$\,mag, respectively.

\section{Individual notes on sample XS-GRBs}

Here we provide notes on each individual burst studied in this work. Since we extracted information about the gas-phase abundance and dust extinction from the literature for all our sample bursts, we will mainly focus on the {\ci} line measurements. For all cases, we only detect a single absorption component from {\ci} but we caution that at this spectral resolution, the observed line profiles might actually be comprised of several narrow lines. However, the fits to H$_2$ for most of the GRBs also indicate that only one component from the molecular-gas phase is present in the spectrum. Any intrinsically narrow absorption components contributing significantly to the observed column densities, would also be present at the location of CO but we do not find any evidence of this. 

By modelling a set of synthetic spectra with imposed {\ci} lines with varying $b$-parameters and $N$(\ci), we estimate that the line profiles are intrinsically saturated for $N$(\ci)$_{\rm tot} \gtrsim 14.5$ (for $b\gtrsim 5$\kms). This is also supported by the typical uncertain broadening parameters associated with the GRBs having the largest {\ci} column densities. For the GRB absorbers where the best-fit value for $N$(\ci) is above this limit (GRBs\,120119A, 150403A, and 180325A), we therefore only provide the $2\sigma$ lower limit on the {\ci} abundances. For the GRB absorption systems with $N$(\ci) < 14.5 (GRBs\,120815A, 121024A, 181020A, and 190114A) we report the measured value for each of the {\ci} abundances. These are also all consistent with the linear relation found for the quasar {\ci} absorbers in Sect.~\ref{sec:sel} (see Fig.~\ref{fig:ciewn}) between $N$(\ci)$_{\rm tot}$ and the rest-frame {\ci} equivalent widths of the same systems. To further verify the robustness of the column density measurements we produce a set of synthetic spectra with varying $b$-parameters and perform the same fitting routine as detailed in Sect.~\ref{ssec:ci}. We are able to recover the input column densities (for non-saturated lines) within the error for all values of $b \gtrsim 2$\kms.

\subsection*{GRB\,120119A at $z=1.7288$} 

The spectrum presented here was published by \cite{Selsing19}. The gas-phase abundances listed in Table~\ref{tab:col} derived for this GRB are adopted from \cite{Wiseman17}. They found $\log N$(H\,{\sc i}) = $22.44\pm 0.12$, [Zn/H] = $-0.96\pm 0.28$ and [Zn/Fe] = $1.04\pm 0.35$. Following \citet{DeCia16} we compute a dust-corrected metallicity, [M/H] = [X/H] - $\delta_X$ (where $\delta_X$ is inferred from the iron-to-zinc depletion), of [M/H] = $0.68\pm0.30$. Both \cite{Japelj15} and \cite{Zafar18b} have derived the visual extinction of this GRB and found a consistent value of $A_V \sim 1$\,mag. We chose to adopt $A_V$ from the latter study (listed in Table~\ref{tab:col}), since they included a full parametrization of the extinction curve in the fit. It was not possible to examine H$_2$ in this absorption system due to the low redshift.

\begin{figure} 
	\centering
	\epsfig{file=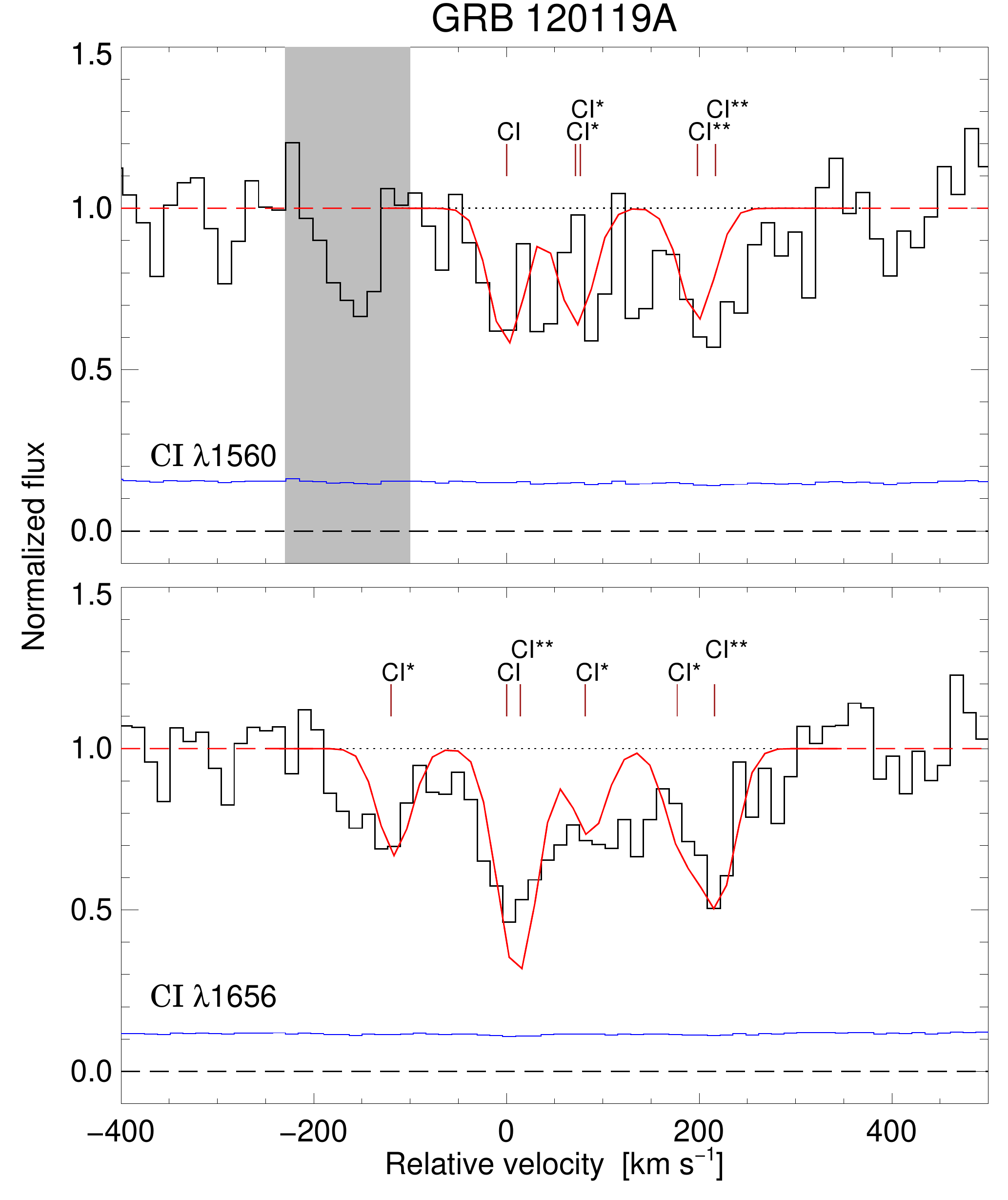,width=9cm}
	\caption{Normalized VLT/X-shooter spectrum of GRB\,120119A in velocity space, centred on the ground-state C\,\textsc{i} transitions at $z=1.72912$. The black solid line shows the spectrum and the associated error is shown in blue. The combined best fit Voigt profiles (with fixed $b=5$\,km\,s$^{-1}$) are indicated by the red solid line, with the red dashed line indicating the continuum. The C\,\textsc{i} ground-state and excited line transitions are marked above each of the absorption profiles. The gray shaded region was ignored in the fit.} 
	\label{afig:120119a_b5}
\end{figure}

To fit the neutral atomic carbon abundances for this GRB we ran two iterations; one where $b$ is left as a free parameter and one where we fix the $b$-parameter to 5\kms. The fit was only constrained by the C\,\textsc{i}\,$\lambda\lambda$\,1560,1656 line transitions since the C\,\textsc{i}\,$\lambda\lambda$\,1277,1328 lines were completely blended with telluric absorption features and located in spectral regions with poor S/N. We also masked out one unrelated line in the immediate continuum region of the C\,\textsc{i}\,$\lambda$\,1560 transition, although it does not appear to be blended with the line profiles. 
We obtain a best-fit $b$-parameter of $b=2.9\pm 0.5$\kms. However, given the S/N of the spectrum we are not able to distinguish which fit is preferred for $b \lesssim 5$\kms. The derived column densities for both $b$-parameters are listed in Table~\ref{atab:120119a}. Since the $b$-parameter cannot be well-constrained we assume $b=5$\kms~and derive a $2\sigma$ lower limit of $\log N$(C\,{\sc i})$_{\rm tot} \gtrsim 14.9$ for this GRB. This lower limit takes into account both the uncertain line broadening and the intrinsically saturated line profiles. The best fit Voigt profiles are shown overplotted on the normalized VLT/X-shooter spectrum in Fig.~\ref{afig:120119a_b5} for $b=5$\kms.

\begin{table}
		\centering
		\caption{Results of the Voigt-profile fitting for GRB\,120119A. The column densities for $b=5$\kms{} are reported throughout.}
		\begin{tabular}{lcc}
			\noalign{\smallskip} \hline \hline \noalign{\smallskip}
			Exc. state & $\log N$(C\,\textsc{i}) \\
			($J$) & $b=2.9\pm 0.5$ & $b=5$  \\
			& (km\,s$^{-1}$) & (km\,s$^{-1}$)  \\
			\noalign {\smallskip} \hline \noalign{\smallskip}
            C\,\textsc{i} & $16.85\pm 0.28$ & $15.10\pm 0.51$ \\ 
            C\,\textsc{i}* & $16.34\pm 0.66$ & $14.16\pm 0.16$ \\ 
            C\,\textsc{i}** & $16.75\pm 0.46$ & $15.13\pm 0.25$ \\ 
			\noalign{\smallskip} \hline \noalign{\smallskip}
		\end{tabular}
		\label{atab:120119a}
\end{table}

\subsection*{GRB\,120327A at $z=2.8143$}

The spectrum presented here was first published by \cite{DElia14}, who also reported the presence of H$_2$ in the spectrum. The neutral and molecular gas-phase abundances listed in Table~\ref{tab:col} derived for this GRB are adopted from \cite{Bolmer19}. They found $\log N$(H\,{\sc i}) = $22.07\pm 0.01$, $\log N$(H$_2$) = $17.39\pm 0.13$, [Zn/H] = $-1.49\pm 0.03$ and [Zn/Fe] = $0.27\pm 0.07$, resulting in a dust-corrected metallicity of [M/H] = $-1.34\pm 0.02$. \cite{DElia14} derived an upper limit on the visual extinction along the line of sight to this GRB of $A_V < 0.03$\,mag. We find no evidence for {\ci} in this absorption system \citep[see also][]{Heintz19a}, and derive a $2\sigma$ upper limit of $\log N$(C\,{\sc i})$_{\rm tot} < 14.3$ assuming $b=2$\kms. 

\subsection*{GRB\,120815A at $z = 2.3582$}

The spectrum presented here was first published by \cite{Kruhler13}, who also reported the presence of H$_2$ and {\ci} in the spectrum. The neutral and molecular gas-phase abundances listed in Table~\ref{tab:col} derived for this GRB are adopted from \cite{Bolmer19}. They found $\log N$(H\,{\sc i}) = $22.09\pm 0.01$, $\log N$(H$_2$) = $20.42\pm 0.08$, [Zn/H] = $-1.45\pm 0.03$ and [Zn/Fe] = $1.01\pm 0.05$, resulting in a dust-corrected metallicity of [M/H] = $-1.23\pm 0.03$. We adopt the visual extinction derived by \cite{Zafar18b} of $A_V = 0.19\pm 0.04$\,mag.

\begin{figure} 
	\centering
	\epsfig{file=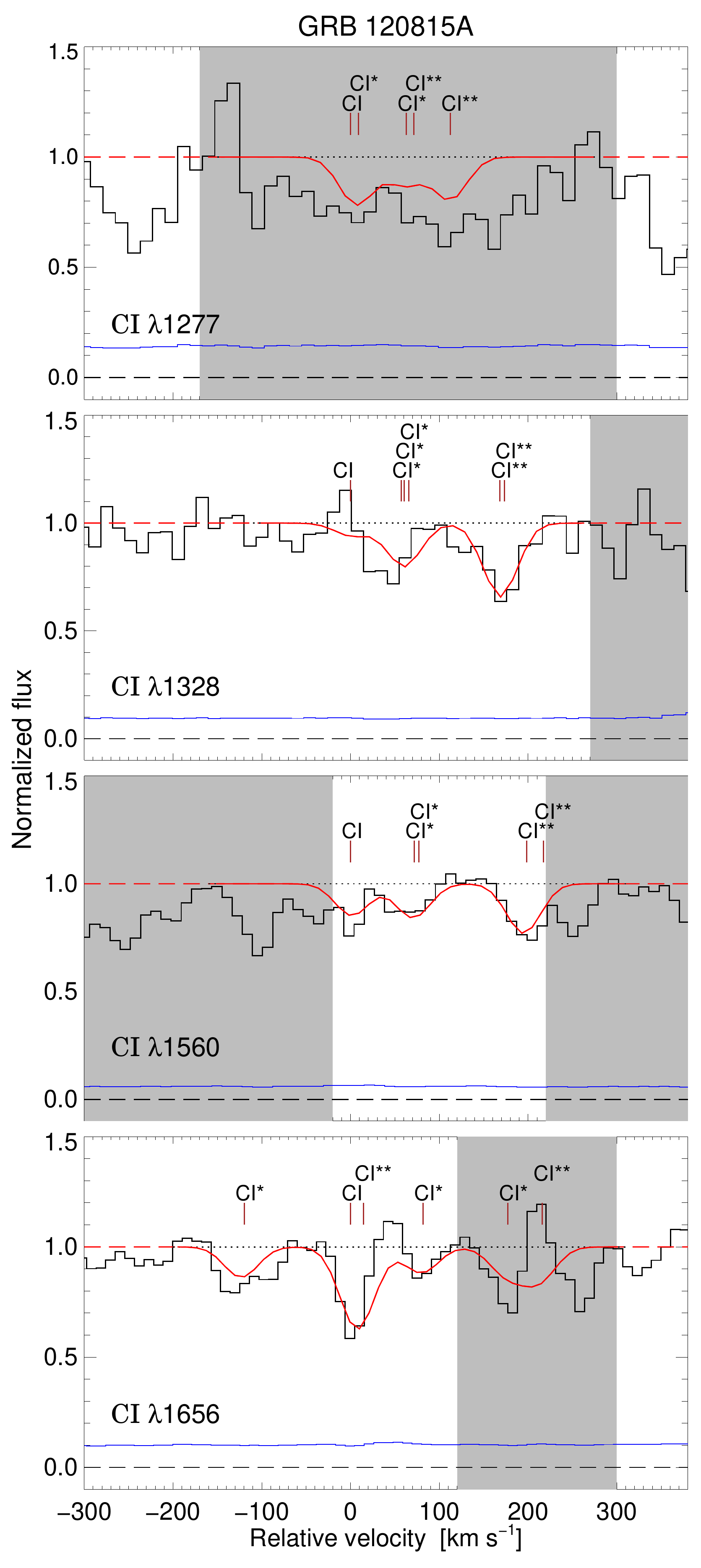,width=9cm}
	\caption{Same as Fig.~\ref{afig:120119a_b5} but for GRB\,120815A with $b=2.3\pm 0.5$, centred on $z=2.35814$.}
	\label{afig:120815a}
\end{figure}

To fit the neutral atomic carbon abundances for this GRB we ran the same two iterations as for GRB\,120119A. Since we only detect a single absorption component from the C\,\textsc{i}\,$\lambda$\,1328 line complex, we fixed this in the fit to the
C\,\textsc{i}\,$\lambda\lambda\lambda$\,1328,1560,1656 line transitions. The C\,\textsc{i}\,$\lambda$\,1277 line was only used to constrain the upper limit on the column density since it is significantly blended with unrelated features. For the line transitions at C\,\textsc{i}\,$\lambda\lambda$\,1560,1656 we also masked out regions of the spectrum showing unrelated absorption features, which were excluded as potential additional velocity components based on the identification in the C\,\textsc{i}\,$\lambda$\,1328 line complex. 
We obtain a best-fit $b$-parameter of $b=2.3\pm 0.5$\kms. The line profiles seem to exclude values of $b\gtrsim 3$\,km\,s$^{-1}$, both when considering the line widths and the relative optical depths. We therefore assume the best-fit $b$ value throughout, but provide the derived column density for fixed $b=5$\kms~in Table~\ref{atab:120815a}. The three fine-structure transitions all show roughly consistent column densities within the errors for both assumed $b$-parameters.
For this GRB we measure a total {\ci} column density of $\log N$(C\,{\sc i})$_{\rm tot} = 14.24\pm 0.14$. Since the lines are not intrinsically saturated this estimate should be reliable, which is also supported by the measured {\ci} rest-frame equivalent width following the linear relation from Fig.~\ref{fig:ciewn}. The best fit Voigt profiles are shown overplotted on the normalized VLT/X-shooter spectrum in Fig.~\ref{afig:120815a} for $b=2.3$\kms.

\begin{table}
		\centering
		\caption{Results of the Voigt-profile fitting for GRB\,120815A.}
		\begin{tabular}{lccc}
			\noalign{\smallskip} \hline \hline \noalign{\smallskip}
			Exc. state & $\log N$(C\,\textsc{i}) \\
			($J$) & $b=2.3\pm 0.5$ & $b=5$ \\ 
			& (km\,s$^{-1}$) & (km\,s$^{-1}$) \\ 
			\noalign {\smallskip} \hline \noalign{\smallskip}
            C\,\textsc{i} & $13.94\pm 0.26$ & $13.35\pm 0.16$ \\ 
            C\,\textsc{i}* & $13.62\pm 0.14$ & $13.52\pm 0.11$ \\ 
            C\,\textsc{i}** & $13.66\pm 0.22$ & $13.57\pm 0.12$ \\ 
			\noalign{\smallskip} \hline \noalign{\smallskip}
		\end{tabular}
		\label{atab:120815a}
\end{table}

\subsection*{GRB\,120909A at $z=3.9290$}

The spectrum presented here was published by \cite{Selsing19}. The neutral and molecular gas-phase abundances listed in Table~\ref{tab:col} derived for this GRB are adopted from \cite{Bolmer19}, who also reported the detection of H$_2$ in the spectrum. They found $\log N$(H\,{\sc i}) = $21.82\pm 0.02$, $\log N$(H$_2$) = $17.25\pm 0.23$, [S/H] = $-1.06\pm 0.12$ and [S/Fe] = $0.50\pm 0.15$, resulting in a dust-corrected metallicity of [M/H] = $-0.29\pm 0.10$. Greiner et al. (in prep.) derived a visual extinction in the line of sight to this GRB of $A_V = 0.16\pm 0.04$\,mag \citep[see][]{Bolmer19}. We find no evidence for {\ci} in this absorption system \citep[see also][]{Heintz19a}, and derive a $2\sigma$ upper limit of $\log N$(C\,{\sc i})$_{\rm tot} < 14.0$ assuming $b=2$\kms. 

\subsection*{GRB\,121024A at $z=2.3005$}

The spectrum presented here was first published by \cite{Friis15}, who also reported the presence of H$_2$ in the spectrum. The neutral and molecular gas-phase abundances listed in Table~\ref{tab:col} derived for this GRB are adopted from \cite{Bolmer19}. They found $\log N$(H\,{\sc i}) = $21.78\pm 0.02$, $\log N$(H$_2$) = $19.90\pm 0.17$, [Zn/H] = $-0.76\pm 0.06$ and [Zn/Fe] = $0.77\pm 0.08$, resulting in a dust-corrected metallicity of [M/H] = $-0.68\pm 0.07$. We adopt the visual extinction derived by \cite{Zafar18b} of $A_V = 0.26\pm 0.07$\,mag.

\begin{figure} 
	\centering
	\epsfig{file=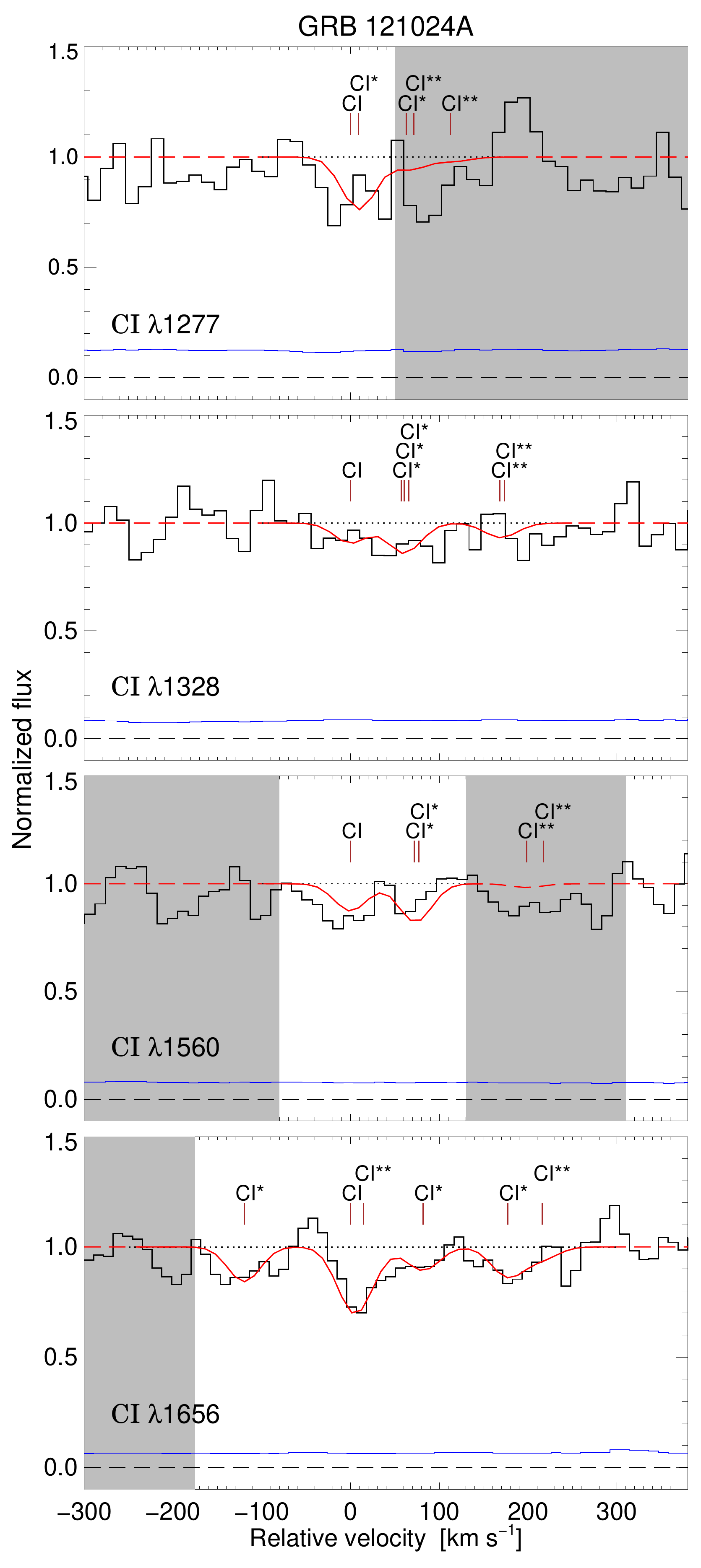,width=9cm}
	\caption{Same as Fig.~\ref{afig:120119a_b5} but for GRB\,121024A with fixed $b=5$\kms, centred on $z=2.30208$.}
	\label{afig:121024a_b5}
\end{figure}

To fit the neutral atomic carbon abundances for this GRB we again ran the fit leaving $b$ as free parameter or fixed to $b=5$\kms. We only detect a single absorption component across the four line complexes so we fixed this in the fit and masked out any unrelated or blended features. The fit was mainly constrained by the C\,\textsc{i}\,$\lambda\lambda$\,1328,1656 line transitions. The C\,\textsc{i}\,$\lambda\lambda$\,1277,1560 lines were only used to constrain the upper limit on the column density since they are significantly blended.
We obtain a best-fit $b$-parameter of $b=3.5\pm 0.5$\kms. The line profiles seem to exclude values of $b\gtrsim 5$\,km\,s$^{-1}$, both when considering the line widths and the relative optical depths. However, since we are not able to distinguish between the best-fit and fixed $b$-parameter of $b=5$\kms, we assume the latter column density throughout. The three fine-structure transitions also all show roughly consistent column densities within the errors for both assumed $b$-parameters.
For this GRB we measure a total {\ci} column density of $\log N$(C\,{\sc i})$_{\rm tot} = 13.91\pm 0.08$. Since the lines are not intrinsically saturated this estimate should be reliable, which is also supported by the measured {\ci} rest-frame equivalent width following the linear relation from Fig.~\ref{fig:ciewn}. The best fit Voigt profiles are shown overplotted on the normalized VLT/X-shooter spectrum in Fig.~\ref{afig:121024a_b5} for $b=5$\kms.

\begin{table}
		\centering
		\caption{Results of the Voigt-profile fitting for GRB\,121024A.}
		\begin{tabular}{lccc}
			\noalign{\smallskip} \hline \hline \noalign{\smallskip}
			Exc. state & $\log N$(C\,\textsc{i}) \\
			& $b=3.5\pm 0.5$ & $b=5$ \\ 
			& (km\,s$^{-1}$) & (km\,s$^{-1}$) \\ 
			\noalign {\smallskip} \hline \noalign{\smallskip}
            C\,\textsc{i} & $13.61\pm 0.17$ & $13.40\pm 0.15$ \\ 
            C\,\textsc{i}* & $13.81\pm 0.10$ & $13.60\pm 0.09$ \\ 
            C\,\textsc{i}** & $13.70\pm 0.17$ & $13.23\pm 0.21$ \\ 
			\noalign{\smallskip} \hline \noalign{\smallskip}
		\end{tabular}
		\label{atab:121024a}
\end{table}

\subsection*{GRB\,141109A at $z=2.9940$}

The spectrum presented here was published by \cite{Selsing19}. The neutral and molecular gas-phase abundances listed in Table~\ref{tab:col} derived for this GRB are adopted from \cite{Bolmer19}, who also reported the detection of H$_2$ in the spectrum. They found $\log N$(H\,{\sc i}) = $22.18\pm 0.02$, $\log N$(H$_2$) = $18.02\pm 0.12$, [Zn/H] = $-1.63\pm 0.06$ and [Zn/Fe] = $0.49\pm 0.07$, resulting in a dust-corrected metallicity of [M/H] = $-1.37\pm 0.05$. \cite{Heintz19a} derived a visual extinction in the line of sight to this GRB of $A_V = 0.11\pm 0.03$\,mag and also found no evidence for {\ci} in this absorption system. We derive a $2\sigma$ upper limit of $\log N$(C\,{\sc i})$_{\rm tot} < 14.7$ assuming $b=2$\kms. 

\subsection*{GRB\,150403A at $z=2.0571$}

The spectrum presented here was published by \cite{Selsing19}. The neutral and molecular gas-phase abundances listed in Table~\ref{tab:col} derived for this GRB are adopted from \cite{Bolmer19}. They found $\log N$(H\,{\sc i}) = $21.73\pm 0.02$, $\log N$(H$_2$) = $19.90\pm 0.14$, [Zn/H] = $-1.04\pm 0.04$ and [Zn/Fe] = $0.63\pm 0.08$, resulting in a dust-corrected metallicity of [M/H] = $-0.92\pm 0.05$. We adopt the upper limit on the visual extinction derived by \cite{Heintz19b} of $A_V < 0.13$\,mag.

\begin{figure} 
	\centering
	\epsfig{file=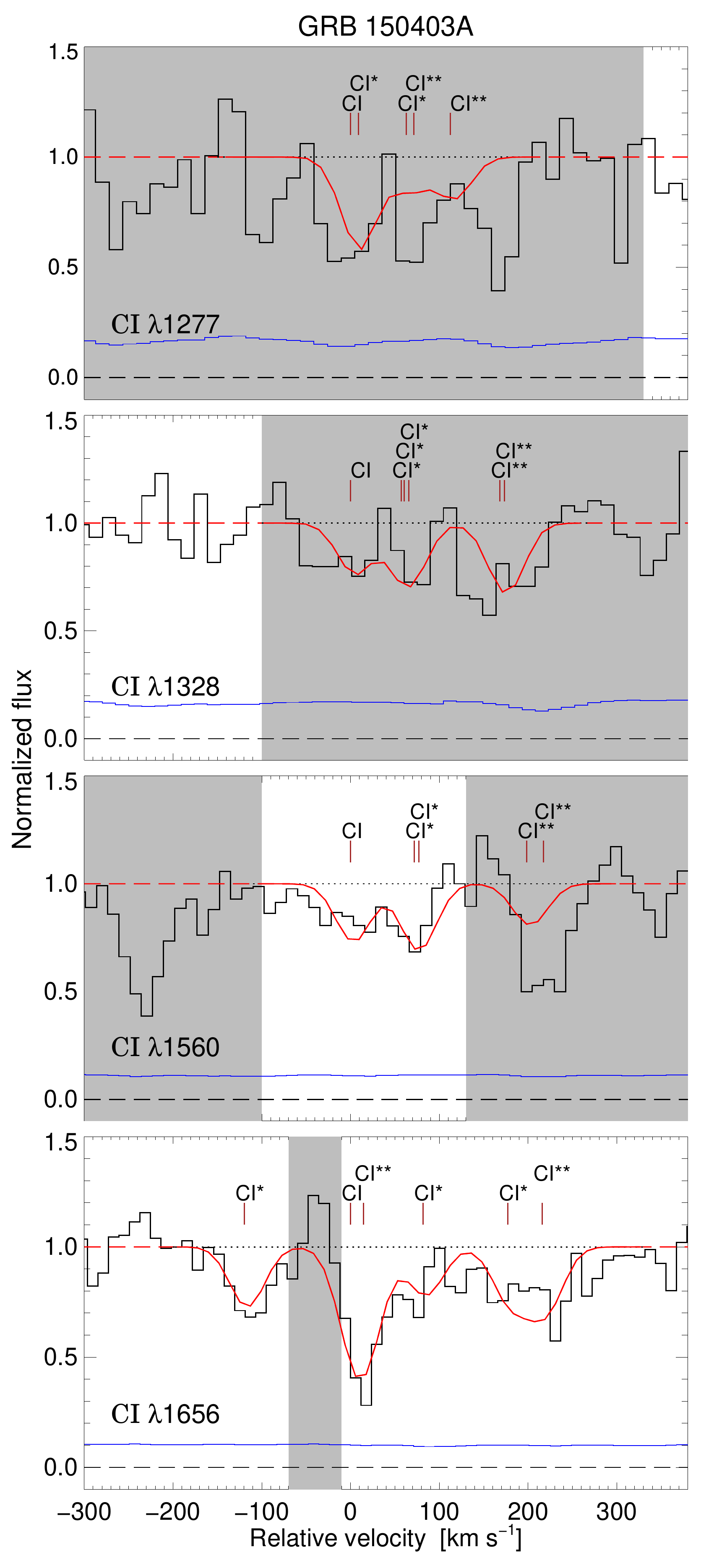,width=9cm}
	\caption{Same as Fig.~\ref{afig:120119a_b5} but for GRB\,150403A with fixed $b=5$\kms, centred on $z=2.05691$.}
	\label{afig:150403a_b5}
\end{figure}

To fit the neutral atomic carbon abundances for this GRB we only ran iterations with fixed values of $b=5$ and $b=10$\kms, since the fit could not converge on a realistic value for $b$ if left as a free parameter. The fit was mainly constrained by the C\,\textsc{i}\,$\lambda\lambda$\,1560,1656 line transitions. The C\,\textsc{i}\,$\lambda\lambda$\,1277,1328 lines were masked out in the fit since they are significantly blended with tellurics. Their apparent optical depths are, however, still required to be consistent with the derived column densities.
For the C\,\textsc{i}\,$\lambda\lambda$\,1560,1656 line transitions we also masked out regions of the spectrum showing unrelated absorption features or bad pixels with correlated noise, and therefore only fit for the same components as identified for the C\,\textsc{i}\,$\lambda$\,1656 line complex. The observed line profiles appear to be consistent with both $b=5$ and $b=10$\kms, so we adopt $b=5$\kms{} to be consistent with the other bursts. The relative abundances for all the fine-structure transitions of the ground-state are consistent, however, within errors for both $b$-parameters. For this GRB we derive a $2\sigma$ lower limit of $\log N$(C\,{\sc i})$_{\rm tot} \gtrsim 14.3$. This lower limit takes into account both the uncertain line broadening and the intrinsically saturated line profiles. The best fit Voigt profiles are shown overplotted on the normalized VLT/X-shooter spectrum in Fig.~\ref{afig:150403a_b5} for $b=5$\kms.

\begin{table}
		\centering
		\caption{Results of the Voigt-profile fitting for GRB\,150403A.}
		\begin{tabular}{lcc}
			\noalign{\smallskip} \hline \hline \noalign{\smallskip}
			Exc. state & $\log N$(C\,\textsc{i}) \\
			& $b=5$ & $b=10$ \\
			& (km\,s$^{-1}$) & (km\,s$^{-1}$) \\
			\noalign {\smallskip} \hline \noalign{\smallskip}
            C\,\textsc{i} & $14.12\pm 0.29$ & $13.74\pm 0.13$ \\
            C\,\textsc{i}* & $14.15\pm 0.15$ & $13.90\pm 0.07$ \\
            C\,\textsc{i}** & $14.50\pm 0.39$ & $14.08\pm 0.12$ \\
			\noalign{\smallskip} \hline \noalign{\smallskip}
		\end{tabular}
		\label{atab:150403a}
\end{table}

\subsection*{GRB\,180325A at $z=2.2496$}

The spectrum presented here was published by \cite{Zafar18a}, who also reported the detection of {\ci}. The gas-phase abundances and visual extinction listed in Table~\ref{tab:col} derived for this GRB are adopted from their work. They found $\log N$(H\,{\sc i}) = $22.30\pm 0.14$, [Zn/H] = $>-0.96$ and $A_V =1.58\pm 0.12$\,mag from the first epoch VLT/X-shooter observations. It was not possible to search for H$_2$ in this absorption system due its high visual extinction.

\begin{figure} 
	\centering
	\epsfig{file=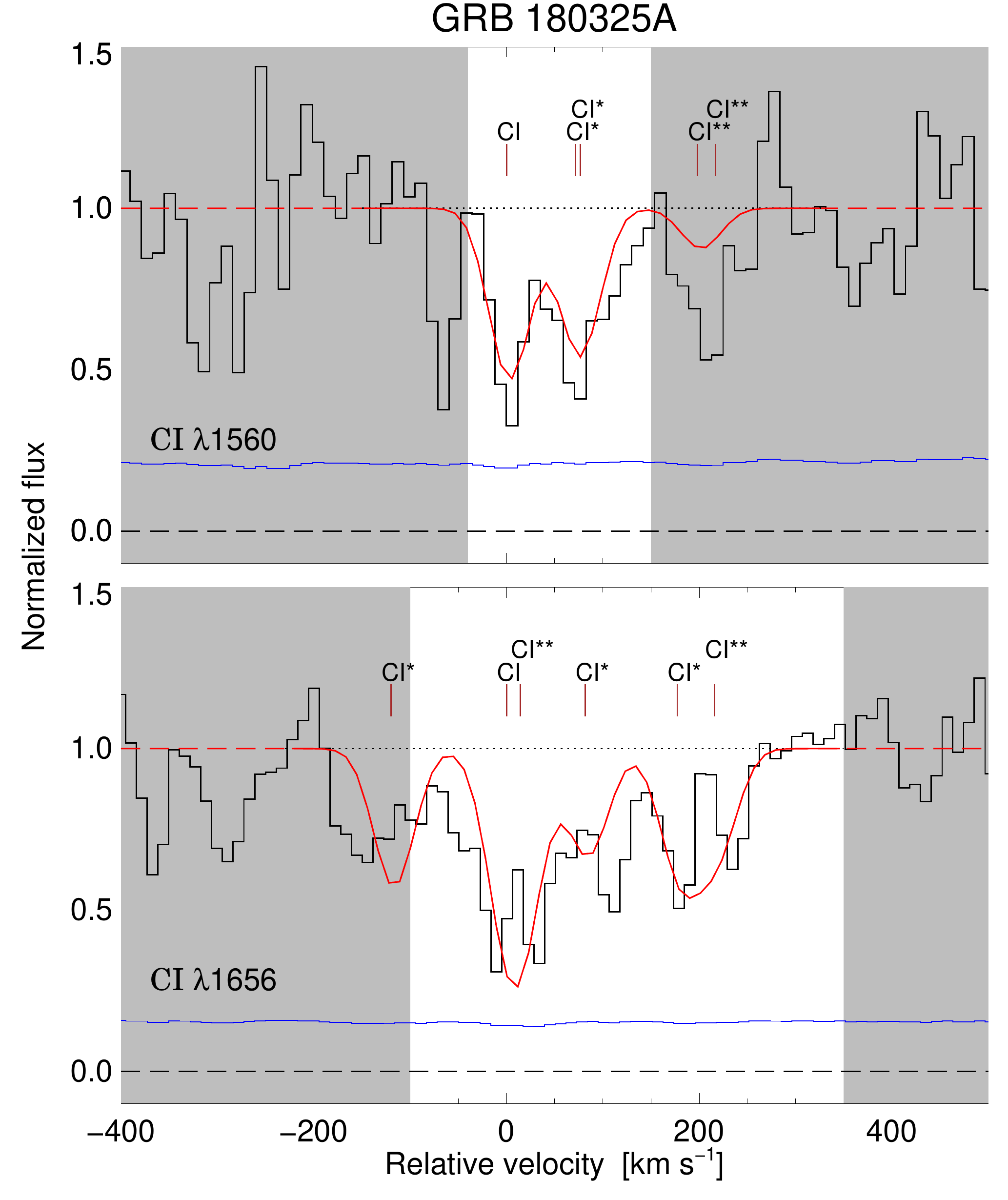,width=9cm}
	\caption{Same as Fig.~\ref{afig:120119a_b5} but for GRB\,180325A with fixed $b=10$\kms, centred on $z=2.24954$.}
	\label{afig:180325a_b10}
\end{figure}

To fit the neutral atomic carbon abundances for this GRB we only ran iterations with fixed $b$-parameters, since the fit could not converge on a realistic value for $b$ if left as a free parameter. The fit was only constrained by the C\,\textsc{i}\,$\lambda\lambda$\,1560,1656 line transitions since the C\,\textsc{i}\,$\lambda\lambda$\,1277,1328 lines were completely blended with telluric absorption features and in spectral regions with poor S/N. For the C\,\textsc{i}\,$\lambda\lambda$\,1560,1656 line transitions we also masked out regions of the spectrum showing unrelated absorption features. The S/N is quite poor in the spectral region around these line complexes as well, so we only perform the fit assuming a single absorption component and with $b$-parameters fixed to $b=5$ and $b=10$\kms, respectively. While the observed line profiles appear to be consistent with both $b=5$ and $b=10$\kms, the column densities derived assuming $b=5$\kms{} (especially the ground-state, see Table~\ref{atab:180325a}) are significantly overestimated. We therefore adopt $b=10$\kms{} and derive a $2\sigma$ lower limit of $\log N$(C\,{\sc i})$_{\rm tot} \gtrsim 14.5$ for this GRB. This lower limit takes into account both the uncertain line broadening and the intrinsically saturated line profiles. The best fit Voigt profiles are shown overplotted on the normalized VLT/X-shooter spectrum in Fig.~\ref{afig:180325a_b10}.

\begin{table}
		\centering
		\caption{Results of the Voigt-profile fitting for GRB\,180325A.}
		\begin{tabular}{lcc}
			\noalign{\smallskip} \hline \hline \noalign{\smallskip}
			Exc. state & $\log N$(C\,\textsc{i}) \\
			& $b=5$ & $b=10$ \\
			& (km\,s$^{-1}$) & (km\,s$^{-1}$) \\
			\noalign {\smallskip} \hline \noalign{\smallskip}
            C\,\textsc{i} & $17.32\pm 0.17$ & $14.50\pm 0.31$ \\
            C\,\textsc{i}* & $14.98\pm 0.27$ & $14.25\pm 0.11$ \\
            C\,\textsc{i}** & $14.51\pm 0.30$ & $14.20\pm 0.20$ \\
			\noalign{\smallskip} \hline \noalign{\smallskip}
		\end{tabular}
		\label{atab:180325a}
\end{table}

\end{document}